\documentclass[draft]{agujournal2019}
\usepackage{url} 
\usepackage{natbib}
\usepackage[inline]{trackchanges} 
\usepackage{soul}
\usepackage{amsmath}
\usepackage{amssymb}
\usepackage[hidelinks]{hyperref} 
\draftfalse

\journalname{Journal of Geophysical Research: Machine Learning and Computation}

\begin{document}

%
%


\title{Electric Potential Patterns Forecasting in the Southern Hemisphere with Deep Learning Techniques}

%
%




\authors{F. P. Ramunno\affil{1, 2}, S. Mestici\affil{3}, I. Coco\affil{4}, M.-T. Walach\affil{5}, S. Massetti\affil{6}, M.F. Marcucci\affil{6}, B. Panos\affil{1}, A. Csillaghy\affil{1}}

\affiliation{1}{University of Applied Sciences and Arts Nortwestern Switzerland}
\affiliation{2}{Department of Computer Science, University of Geneva, 1211 Geneva, Switzerland}
\affiliation{3}{Universit\'a Roma "La Sapienza"}
\affiliation{4}{Istituto Nazionale di Geofisica e Vulcanologia, Via di Vigna Murata, 605, Rome, Italy}
\affiliation{5}{School of Physics and Astronomy, Lancaster University, Lancaster, UK}
\affiliation{6}{Italian National Institute of Astrophysics - Institute for Space Astrophysics and Planetology, Via del Fosso del Cavaliere, 100, Rome, Italy}




\correspondingauthor{Francesco Pio Ramunno}{francesco.ramunno@fhnw.ch}



\begin{keypoints}
\item We release an open, paired dataset of five years of Southern Hemisphere SuperDARN maps matched to L1 solar wind data from 2020 to 2025
\item Conditioning on upstream L1 solar wind data is essential for accurate ionospheric forecasts, especially during storms
\item The diffusion model generalizes best on an out-of-distribution 2015 storm, using L1 data from a different satellite than training
\end{keypoints}

%
%

%
%


\begin{abstract}
Space weather disturbances driven by the solar wind can degrade satellite navigation, disrupt radio communications, and threaten power infrastructure, making accurate forecasting of the high-latitude ionosphere's response a critical operational need. Existing approaches, empirical climatological models and physics-based magnetohydrodynamic simulations, either smooth out the ionosphere's time-dependent, non-linear response or are too computationally expensive for real-time use, while prior machine learning efforts have mostly targeted scalar indices rather than the full spatial structure of ionospheric convection. Here we train and compare three deep learning architectures, a probabilistic diffusion model conditioned on multi-variate solar wind and interplanetary magnetic field measurements at the L1 Lagrange point, an unconditioned diffusion ablation, and a deterministic U-Net baseline, to forecast Southern Hemisphere high-latitude electric potential maps derived from SuperDARN radar observations. Using five years (2020--2025) of SuperDARN data synchronised with DSCOVR L1 measurements, we evaluate the models under a single-pass regime, an extended autoregressive rollout of up to 350 frames, and an out-of-distribution case study on the intense March 2015 St.\ Patrick's Day storm, unseen during training. We find that the relative advantage of the deterministic and probabilistic approaches is not fixed: the deterministic model is competitive over short horizons and calm conditions, while the diffusion model's advantage grows and eventually dominates as the forecast horizon lengthens and the event becomes more dynamic, evidence that probabilistic, sample-based generative models are the more promising direction for operational, long-horizon space weather forecasting.
\end{abstract}

\section*{Plain Language Summary}
The Sun releases a constant stream of charged particles, the solar wind, which can disturb the electrically charged upper atmosphere above Earth's poles, interfering with GPS, radio communication, and power grids. We trained three AI systems to predict how these disturbances evolve, using five years of radar data from Antarctica alongside real-time solar wind measurements from a satellite between the Sun and Earth. One system generates several possible outcomes for each forecast instead of a single guess, capturing the uncertainty in what might actually happen. We tested all three on both routine conditions and an intense storm from 2015 that none had seen before. The simpler, single-guess system did fine for short, calm periods, but the system generating multiple outcomes became clearly better the further ahead it predicted and the more disturbed was the ionospheric environment, making it the more promising choice for forecasting the most disruptive space weather events.

%
%

%


%
%
%
%

\section{Introduction}

The increasing reliance of modern society on space- and ground-based technological infrastructure has transformed the domain of space weather from a specialised field of heliophysics to a critical operational priority. The Earth's Magnetosphere-Ionosphere-Thermosphere (M-I-T) system is a complex, coupled environment that responds dynamically to solar forcing \citep{milan2017}. Disturbances within this system can severely degrade the precision of Global Navigation Satellite Systems (GNSS), disrupt trans-ionospheric radio communications, and reduce satellite operational lifetimes through enhanced atmospheric drag. Large-scale geomagnetic storms, typically triggered by the arrival of Coronal Mass Ejections (CMEs) or sustained periods of southward Interplanetary Magnetic Field (IMF), can induce powerful electric currents (Geomagnetically Induced Currents, or GICs) in terrestrial power grids and pipelines, posing a direct threat to national infrastructure and energy security \citep[e.g][]{smith2024}. Recent severe geomagnetic storms have starkly illustrated these risks, reinforcing the necessity of developing risk-mitigating services supported by robust forecasting capabilities. For the machine learning community, this environment represents a formidable challenge with a high-dimensional, non-linear system characterised by multi-scale interactions and chronically sparse observational data.

Central to mitigating these risks is the characterisation of the Earth's Ionosphere, a multi-layered partially ionised region of the upper atmosphere extending from an altitude of 80 up to a thousand kilometers, which serves as the immediate environment for many technological systems. In the high-latitude regions, the ionosphere acts as a dissipative load within a global circuit, converting electromagnetic energy transferred from the solar wind and magnetosphere into thermospheric heat via Joule dissipation. Electrical coupling between these regions is maintained along geomagnetic field lines, which are generally assumed to be electric equipotentials. In this framework, energy transfer and dissipation is realised though large-scale horizontal plasma convection driven by $\textbf{E} \times \textbf{B}$ drift, where $\textbf{E}$ represents the electric field transferred by the solar wind into the ionosphere and $\textbf{B}$ is the geomagnetic field. The spatial distribution of plasma convection at high latitudes can be characterised in terms of an electrostatic potential ($\phi$) where $\mathbf{E} = -\nabla\phi$.

The most significant energy injection occurs during periods of southward IMF (B$_z<$0), where magnetic reconnection is triggered on the dayside magnetopause \citep{dungey1961}, opening geomagnetic field lines and allowing the solar wind to map directly into the polar cap. This process drives a characteristic double cell convection pattern, featuring anti-sunward flow across the central polar cap and return sunward flow at lower latitudes (up to $\sim$50° magnetic Latitude) within the auroral regions. The symmetry of this system is controlled by the IMF B$_y$ component, which leads to more complex geometries resulting in one dominant circular cell and one crescent shaped cell \citep{cowley1981}. In the Northern Hemisphere, positive $B_y$ shifts the flow toward the dawn side, while negative $B_y$ biases it towards dusk and these dependencies appear in the opposite sense in the Southern Hemisphere. Conversely, for northward IMF ($B_z > 0$), the lack of dayside reconnection between open and closed magnetic field lines means remnant convection is driven by nightside reconnection, cusp reconnection on open field lines and the viscous interaction between the solar wind and the magnetopause. Under these conditions, convection is constrained to high latitudes ($>$70° magnetic Latitude) and manifests as more complex patterns, often involving up to four cells with a signature sunward flow component in the central polar cap \citep{reiffburch1985}.

Despite this foundational understanding, the described system remains a simplified representation. Indeed, this description does not fully account for the injection of energy from the magnetotail on the nightside, where magnetic reconnection releases intermittent plasma flows toward the auroral regions. These nightside processes are characterised by longer timescales and a time-delayed response relative to dayside forcing, leading to highly non-linear interactions. Furthermore, due to the intrinsic inertia in ionospheric dynamics and the non-linear coupling between its components, the development and triggering mechanisms of smaller-scale patterns remain elusive. This inherent complexity has necessitated decades of observation using a multi-instrument approach involving satellites, balloons, rockets, and ground-based radar networks. In particular, the Super Dual Auroral Radar Network (SuperDARN) has fundamentally shaped our understanding of how the ionosphere responds to IMF orientation, and solar wind speed and pressure \citep{greenwald1995,nishitani2019} by providing global-scale albeit sparse, high-resolution measurements of plasma drift velocities across both hemispheres.

Despite the wealth of observational data, accurately forecasting the evolution of ionospheric convection remains a challenge. Current modelling efforts primarily fall into two categories: empirical/climatological models and global Magnetohydrodynamic (MHD) simulations. Empirical models \citep{weimer2005,cousins2010,thomas2018,lam2023,walach2025} provide robust global descriptions of the average ionospheric convection patterns associated with different IMF conditions. However, these models inherently smooth out mesoscale features and struggle to capture the non-linear, time-dependent response of the ionosphere to transient solar wind forcing. Conversely, while physics-based MHD models provide a first-principles approach to geospace dynamics \citep{toth2005swmf}, they are computationally expensive, may require multi-source input data not available in an operational forecasting context, and are often limited in their ability to resolve the complex boundary layers at the ionospheric interface in real time. The emergence of data-driven methodologies offers a promising path forward, yet machine learning has so far been applied mainly to scalar space weather indices (e.g., Dst or Kp prediction); its application to the high-dimensional spatial problem of global convection mapping is still growing, as we discuss in Section~\ref{sec:related_works}.

In this paper, we train and compare three deep learning architectures, a probabilistic diffusion model conditioned on multi-variate solar wind and IMF drivers, an unconditioned diffusion ablation, and a deterministic U-Net baseline, to forecast the spatial structure of high-latitude ionospheric electric potential maps. Leveraging a comprehensive five-year dataset (2020--2025) of Southern Hemisphere SuperDARN observations, synchronised with high-resolution L1 measurements from the DSCOVR mission and constrained by the \citet{cousins2010} climatological background, we evaluate these models under progressively more demanding conditions: a single-pass regime matching the training setup, an extended autoregressive rollout of up to 350 predicted frames, and a fully out-of-distribution case study on an intense Solar Cycle 24 storm unseen by any of the models during training. We show that the relative merit of the deterministic and probabilistic approaches is not fixed, but shifts systematically with the length of the forecast horizon and the true dynamism of the convection pattern.

\section{Related Works}
\label{sec:related_works}

In this section, we discuss our work relative to prior machine learning applications to space weather and, specifically, SuperDARN convection mapping.

\subsection{Machine Learning for Space Weather}
\label{subsec:ml_space_weather}

Machine learning has been applied across the space weather pipeline, from solar activity forecasting to
geomagnetic index prediction. \citet{camporeale2019} provides a comprehensive review of this growing body
of work, framing nowcasting and forecasting as the two dominant paradigms and highlighting neural
networks' capacity to learn non-linear input-output mappings directly from data, in contrast to the
linear or climatological relationships underlying earlier empirical models. Within this landscape, most
effort has concentrated on scalar or low-dimensional targets, such as geomagnetic indices like Dst and
Kp, relativistic electron flux, or solar wind speed at 1~AU, where a single time series is predicted from
a small set of upstream drivers \citep[][and references therein]{cristoforetti2022}.

Extending these approaches to the ionosphere introduces a substantially harder, spatially-resolved
prediction problem: rather than a scalar index, the target is a two-dimensional field (total electron
content, critical frequency, or, in our case, the electric potential) that varies simultaneously in space
and time. Deep learning applications to ionospheric TEC forecasting have shown that convolutional and
recurrent architectures can capture the spatial structure and storm-time evolution of these fields more
accurately than empirical climatologies \citep{ren2024tec,mestici2025ioncast}, but the high-latitude regions
specifically remains comparatively underexplored, due to their more complex dynamics and sparser observations.

Direct application of deep learning to SuperDARN-derived convection products is correspondingly more
recent and still limited in number. \citet{liu2020cpcp} were, to our knowledge, among the first to apply
deep learning to a SuperDARN-derived target, using a multi-layer perceptron and an Long Short-Term Memory (LSTM) network to
estimate the scalar Cross Polar Cap Potential from solar wind and IMF parameters, improving Root Mean Square Error (RMSE) from
above 10~kV for prior empirical fits to 7.20~kV. \citet{deng2022superdarn} extended this to the full
spatially-resolved convection potential map, comparing a plain feedforward network, an FC-LSTM, and an
encoder-decoder ConvLSTM, and found that only the spatiotemporal ConvLSTM architecture, which explicitly
models both spatial structure and temporal history, reproduced the cross polar cap potential distribution
well; the purely feed-forward and non-convolutional recurrent baselines could not fully capture the
spatial structure of the field. Most recently, \citet{gao2026superdarn} trained a family of multi-layer
perceptron architectures (a plain MLP, an MLP augmented with multi-head attention, and a residual MLP) to
map six instantaneous near-Earth parameters, five solar wind/IMF measurements (Bx, By, Bz, Vx, Pd) and the
geomagnetic AE index, directly onto the high-latitude convection pattern,
reporting that their best configuration (Res-MLP) reaches a structural similarity index of 0.54 and a
correlation of 0.88 against SuperDARN-derived maps. The inputs used are not purely independent of the upstream driver because AE is itself derived from ground-based
auroral-zone magnetometers measuring the ionospheric electrojet currents and is a signature of the same
magnetosphere-ionosphere coupling the model is trained to predict. Accordingly, the task is closer to a same-time regression between two ionosphere-linked observables than a forecast
driven solely by upstream forcing. This mismatches their explicit design goal, which is rapid inference from a
minimal set of instantaneous input parameters with minimal reliance on historical observations, targeting
nowcasting speed rather than a probabilistic characterisation of the convection pattern or an assessment
of forecast stability over an extended rollout.

Our work is positioned to address several gaps this line of research leaves open. First, all of the
architectures above, including \citet{gao2026superdarn}'s, are deterministic point estimators: they
return a single most-likely map rather than a distribution over plausible convection patterns, precluding
any notion of forecast uncertainty. We instead adopt a diffusion-based generative formulation
(Section~\ref{subsec:diffusion}), which produces an ensemble of physically plausible realisations per
forecast and, as we show in Section~\ref{subsec:results_singlepass}, is not merely a modelling preference
but changes which model is preferable as the dynamism of the target sequence increases. Second,
single-realisation structural metrics such as Structural Similarity Index Measure (SSIM), the headline metric reported by
\citet{gao2026superdarn}, are, by construction, biased toward smoothed, deterministic outputs
\citep{saharia2022,dahl2017,ledig2017,menon2020}. Our evaluation reports SSIM alongside NRMSE of physically motivated scalar statistics, a power-spectrum-based metric, and Learned Perceptual Image Patch Similarity (LPIPS) specifically to expose
this bias rather than be misled by it (Section~\ref{subsec:metrics}). Third, and most fundamentally,
prior SuperDARN deep learning models have been validated on data drawn from the same distribution and,
implicitly, the same short forecast horizon as their training set. We instead evaluate stability under a
much longer autoregressive rollout (Section~\ref{subsec:results_fakear}) and, critically, under the fully
out-of-distribution storm case study already introduced above (Section~\ref{subsec:results_2015}). Finally, our conditioning is restricted only to upstream L1 solar wind
and IMF measurements (Section~\ref{subsec:l1_data}), therefore preserving the forecast character of the task.

\section{Data}

The data used to train and validate the three models in this work considers an interval from 2020 to 2025, covering the ascending phase of the 25th solar cycle from minimum to maximum and thus a wide range of geomagnetic conditions. We also include as a case study March 2015, which was characterised by high geomagnetic activity during the peak of the 24th solar cycle. We first describe the physical preprocessing pipeline that produces the ionospheric potential maps and associated solar wind/geomagnetic context (Sections~\ref{subsec:l1_data}--\ref{subsec:indices}), and then the machine-learning-specific preprocessing applied before the data are presented to the models (Section~\ref{subsec:ml_preprocessing}).

\subsection{Solar Wind and Interplanetary Magnetic Field Data}
\label{subsec:l1_data}

We first consider upstream solar wind and IMF measurements, which represent the primary drivers of energy and momentum transfer into the magnetosphere-ionosphere system. For the longer 2020-2025 interval, we use 1-minute cadence data from the Deep Space Climate Observatory (DSCOVR) mission \citep{BJS2012} in Geocentric Solar Ecliptic (GSE) coordinates. For the March 2015 case study, we instead rely on 1-minute measurements from the Advanced Composition Explorer (ACE) mission \citep{stone1998ace}. Both spacecraft are located at the L1 Lagrangian point and provide continuous, high-resolution solar wind plasma and IMF measurements.

For the entire time interval under study we also use one-minute resolution OMNI data \citep{king2005omni}, collected from the CDAWeb database\footnote{\url{https://omniweb.gsfc.nasa.gov}}. The OMNI variables are provided in Geocentric Solar Magnetospheric (GSM) coordinates and are already propagated from L1 to the bow shock nose using the bow shock model of \citet{farris1994bowshock}. Unlike DSCOVR, OMNI is not used as a model input: it serves exclusively for the CS10 fitting step described in Section~\ref{subsec:superdarn}.

\begin{figure}
    \centering
    \includegraphics[width=\linewidth]{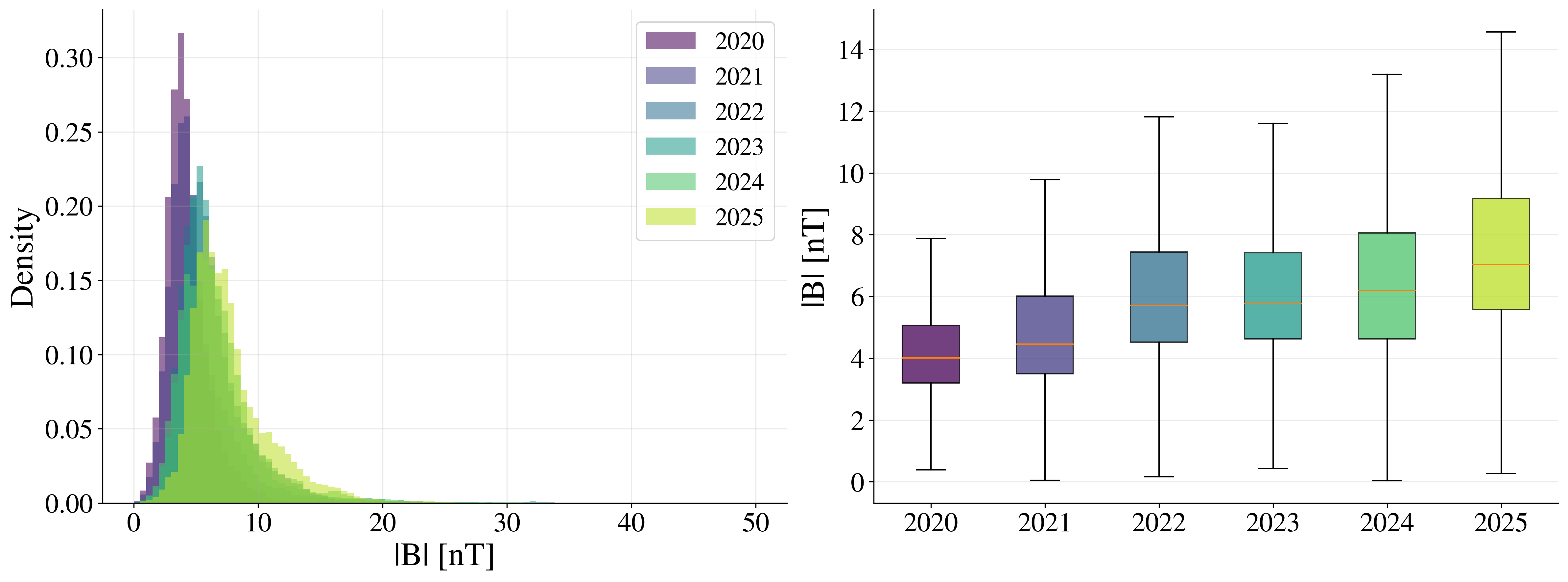}
    \caption{Distribution of the total interplanetary magnetic field magnitude $|B|$ per year, 2020--2025
    (histogram, left; boxplot without outliers, right). Both the typical magnitude and its spread
    increase monotonically from 2020 to 2025, reflecting the ascending phase of Solar Cycle 25 covered by
    the training and validation data.}
    \label{fig:b_dist_per_year}
\end{figure}

Figure~\ref{fig:b_dist_per_year} shows the resulting distribution of the total IMF magnitude $|B|$ across the 2020--2025 interval, year by year: both its typical value and its spread increase monotonically, reflecting the ascending phase of Solar Cycle 25 spanned by the training and validation data and, correspondingly, the widening range of driving conditions the models are exposed to over the course of the dataset.

\subsection{SuperDARN Observations and Electric Potential Mapping}
\label{subsec:superdarn}

Ionospheric plasma drift velocities are obtained from the Super Dual Auroral Radar Network (SuperDARN), an international global array of high-frequency (HF) coherent-scatter radars spanning high- and mid-latitude regions in both hemispheres \citep{greenwald1995,nishitani2019}. In this study, we focus exclusively on observations from the Southern Hemisphere, a scientifically challenging region owing to its complex environment and sparser radar distribution, making it a critical test of model robustness under less favorable data coverage than the Northern Hemisphere.

Individual SuperDARN radars measure the line-of-sight (l-o-s) component of the bulk $E \times B$ drift of ionospheric plasma, through the Doppler phase shift of coherent scatter of HF radio pulses against plasma irregularities  drawn into the large-scale drift.
The raw radar data are processed using the Radar Software Toolkit (RSTv5.1) package \citep{thomas2025}. We utilise the FitACF v3.0 routine to derive Doppler velocities from the original autocorrelation functions (ACFs). This version includes refined ACF fitting packages and standardises the removal of ground scatter and or noise. Following standard SuperDARN processing techniques, the l-o-s velocity data are first mapped onto a spatial grid with a 1° magnetic latitude and longitude resolution at a 2 minutes temporal cadence. This mapping procedure is performed in a Magnetic Local Time (MLT) and Altitude Adjusted Corrected Geomagnetic Latitude (AACGM, \citet{Baker1989}) reference frame. 


We compute the instantaneous ionospheric convection patterns using the electrostatic potential ($\Phi$) map technique \citep{ruohoniemi1998}, by fitting the observed l-o-s velocities to a spherical harmonic expansion. In order to constrain the fitting procedure in regions where radar backscatter is absent or very poor, we supplement the observed data with model data sampled from the \citet{cousins2010} (CS10) climatological model. We use CS10 rather than higher-resolution alternatives such as TS18 \citep{thomas2018} because it was explicitly derived using also Southern Hemisphere observations, ensuring the resulting potential maps are physically consistent with southern hemispheric dynamics.

The CS10 model is driven by the IMF, the solar wind electric field ($E_{sw}$) and the Earth’s dipole tilt angle. We achieve temporal alignment by assigning the respective IMF and plasma parameters from the 1-minute OMNI dataset to each 2-minute SuperDARN integration period. For the fitting process, we adopt a spherical harmonic expansion of order 8. For the gridding process, we set the Heppner-Maynard Boundary (HMB) \citep{heppnermaynard1987}, which defines the lower-latitude limit of the convection zone, to a maximum of -65° AACGM (thus covering the region from -65° to -89° AACGM). This fitting procedure generates spherical harmonic coefficients of the best fit and therefore allow us to generate a 2D potential map spanning 24° in AACGM and 360° in MLT with a 1° resolution for both dimension. For the purposes of this study, we generate electrostatic potentials corresponding to a matrix with a (24, 360) shape for each time instance.

\subsection{Solar and Geomagnetic Activity Indices}
\label{subsec:indices}

To characterise the prevailing heliophysical conditions and the corresponding magnetospheric response during the study period, we make use of several standard solar and geomagnetic indices. Solar activity levels are monitored using the F10.7 cm radio flux and the International Sunspot Number (SSN), proxies for solar extreme ultraviolet (EUV) radiation and the phase of the solar cycle, respectively.


Finally, to quantify the potential energy transfer from the solar wind into the magnetosphere, we derive the Perreault-Akasofu Epsilon ($\epsilon$) parameter \citep{akasofu1981}. This coupling function is calculated using the IMF magnitude ($B$), solar wind velocity ($v$), and the IMF clock angle ($\theta$) obtained from the DSCOVR measurements:$$\epsilon = \frac{4\pi}{\mu_0} v B^2 \sin^4\left(\frac{\theta}{2}\right) l_0^2$$where $l_0$ is a characteristic length scale (typically $7 R_E$). Collectively, these indices and parameters provide a multi-scale description of the solar-terrestrial environment, allowing the visualisation of various levels of solar and geomagnetic disturbances.

\subsection{Machine Learning Preprocessing}
\label{subsec:ml_preprocessing}

Before being presented to the models, the (24, 360) AACGM/MLT potential map of Section~\ref{subsec:superdarn} undergoes a further, purely machine-learning-motivated preprocessing step, distinct from the physical preprocessing described above. First, the electrostatic potential is resampled onto a $128\times128$ Cartesian pixel grid centred on the magnetic pole via linear interpolation, with the -66 to -90° AACGM annulus inscribed as a disk and the four corners of the square outside that disk zero-padded. This Cartesian representation, rather than the native polar grid, is what is passed to the convolutional and attention-based backbones described in Section~\ref{sec:background}. Second, the electric potential is clamped to $\pm 80{,}000$~V and linearly rescaled to $[-1, 1]$, matching the input range expected by the diffusion and U-Net architectures. This bound was set by inspecting the empirical pixel-level distribution of the electric potential across the full 2024 subset of the dataset (263,152 maps): 99.99\% of pixels fall below $\pm53{,}900$~V, and fewer than $10^{-4}\%$ of pixels exceed the $\pm80{,}000$~V bound, so the clamp discards a negligible fraction of the dynamic range while keeping the normalisation to $[-1, 1]$ well-conditioned for stable training.

\begin{figure}
    \centering
    \includegraphics[width=\linewidth]{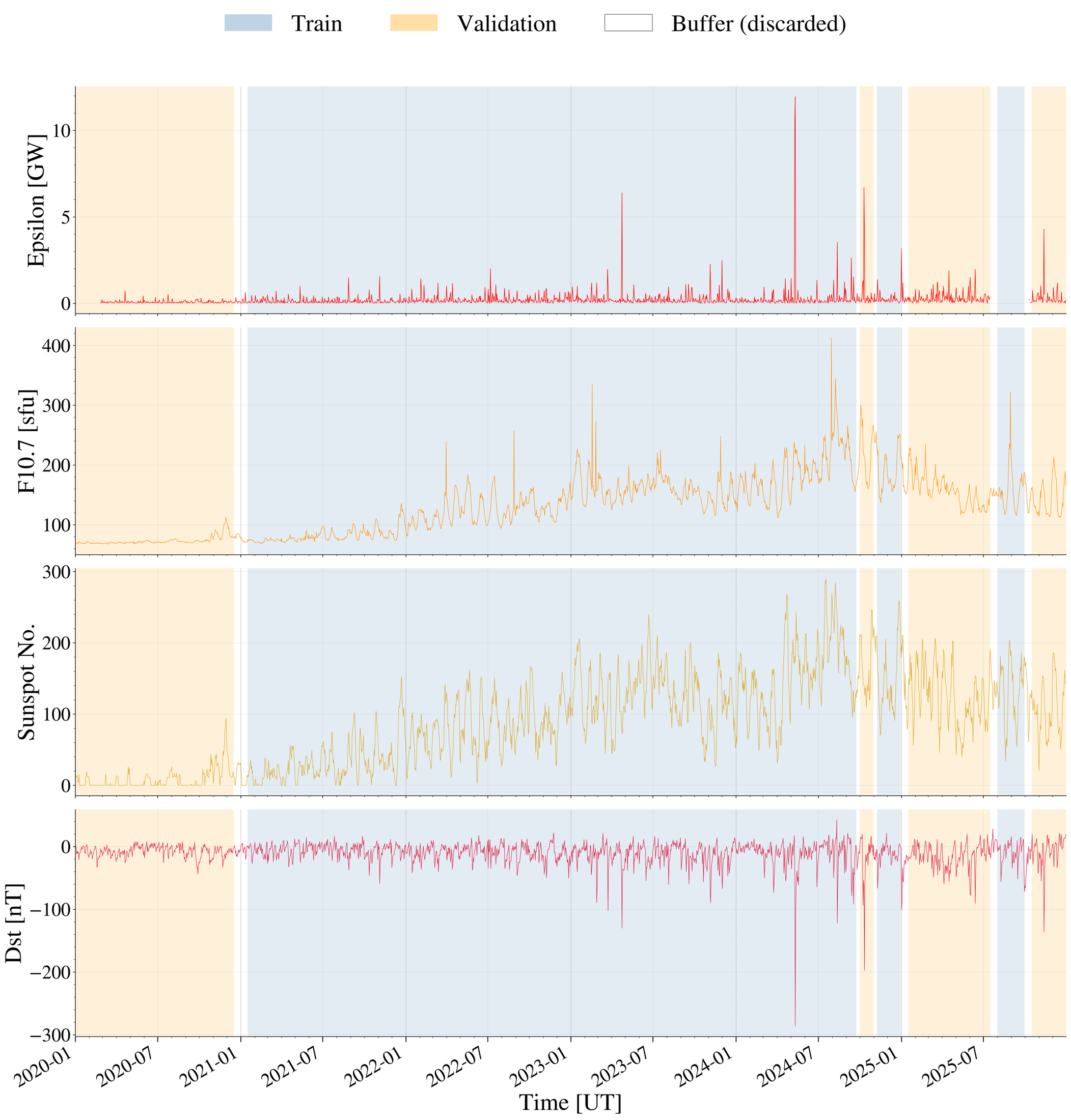}
    \caption{Chronological train/validation split, 2020--2025, shown against the Akasofu epsilon coupling
    parameter and the F10.7, sunspot number, and Dst indices (Section~\ref{subsec:indices}). Blue
    shading marks training regions, orange marks validation regions, and unshaded (white) intervals are
    15-day buffer zones discarded from both splits to prevent data leakage between adjacent regions.}
    \label{fig:train_val_split}
\end{figure}

Overlapping input/target sequences of $N+M=22$ frames, each frame pairing one electric potential map with the temporally aligned L1 solar wind/IMF vector (Section~\ref{subsec:data_alignment}), are built from the continuous 2020--2025 record by sliding a window across the merged L1/SuperDARN timeseries (44 minutes at 2-minute cadence: $N=15$ conditioning frames and $M=7$ target frames, formally defined in Section~\ref{subsec:problem_formulation}),
retaining only sequences with no missing data. To limit redundancy between adjacent, heavily overlapping windows, we do not retain every possible window position: consecutive retained sequences must have their $t_0$ (Section~\ref{subsec:problem_formulation}) at least 15 frames (30 minutes) apart in the training split and at least 30 frames (60 minutes) apart in the validation split.
Each sequence is then assigned to training or validation based on the calendar period its $t_0$ falls into (several separate date ranges across 2020--2025, not a single early/late split).
Figure~\ref{fig:train_val_split}  shows the intervals corresponding to training (pale-blue shaded areas) and validation (pale-orange shaded areas) sets throughout our dataset; each panel shows the behaviour of the major solar and geomagnetic indices (see Section~\ref{subsec:indices}), from top to bottom: the Perreault-Akasofu Epsilon ($\epsilon$) parameter, the F10.7 solar index, the Sunspot number and the Dst index. As clearly emerges from the Figure, the training and validation regions are also interleaved across the full 2020--2025 span rather than each being a single contiguous block, so both splits sample a similar range of geomagnetic activity, from quiet to storm-time, rather than one being biased toward the quieter early years and the other toward the more active later ones. This matters for the validation metrics of Section~\ref{sec:results} to be meaningful: an active-regime-poor validation split would make strong validation performance reflect only skill on calm conditions, not on the dynamic, high-activity regimes that matter most operationally.
Sequences whose $t_0$ falls in neither region are discarded as a 15-day buffer rather than assigned to either split: since adjacent sequences overlap in time, a sequence immediately adjacent to a boundary would otherwise share most of its frames, and thus near-identical data, with one on the other side of it. The buffer removes this overlap entirely, keeping the validation set an independent,
data-leakage-free test of generalisation rather than a near-copy of training data.

In total, the training set contains 43,781 sequences and the validation set contains 9,168 sequences.

\subsection{Solar Wind/Interplanetary Magnetic Field and Ionosphere Data Alignment}
\label{subsec:data_alignment}

Because the L1 measurements and the SuperDARN electric potential maps are recorded by different instruments at different physical locations, pairing them into a single training sequence requires accounting for the finite time it takes the solar wind to propagate from L1 to Earth, rather than treating the two modalities as simultaneous. As described in Section \ref{subsec:l1_data}, the L1 parameters are obtained by the DSCOVR satellite, so we first compute the distance between the DSCOVR satellite and Earth and then use it to compute the solar wind Earth arrival time by using its speed, aligning each L1 measurement with the corresponding electric potential map. At this stage we do not consider the physical problem of faster solar wind structures that may overtake and absorb slower structures, or the time taken for magnetopause-magnetosphere transition for any solar wind structures to take effect but we will consider this more complex aspect in future work. Thus, for simplicity, we have aligned each electric potential map with one solar wind data point observed by DSCOVR.

\section{Background}
\label{sec:background}

In this section, we describe the technical background for the architectures compared in this work: the diffusion-based generative framework drawing on \citet{Karras2022}, the Vision Transformer on \citet{dosovitskiy2020}, and the U-Net on \citet{ronneberger2015}.

\subsection{Diffusion-Based Generative Models}
\label{subsec:diffusion}

Diffusion models \citep{sohldick} define a generative framework through a forward process that gradually corrupts data with Gaussian noise and a learned reverse process that recovers the data distribution, developed along two main lines: Denoising Diffusion Probabilistic Models (DDPM) \citep{ho2020} and score-based generative models via stochastic differential equations \citep{song2019, song2021sde}. \citet{Karras2022} unify and reparametrise these into the Elucidated Diffusion Model (EDM), which we adopt in this work, systematically analysing their shared design space and replacing discrete variance schedules with a continuous noise-level parametrisation.

In the EDM framework, the forward process is defined by a perturbation kernel that corrupts a clean sample $x_0$ by adding Gaussian noise with standard deviation $\sigma$:
\begin{equation}
    p(\tilde{x} | x_0) = \mathcal{N}(\tilde{x};\, x_0,\, \sigma^2 \mathbf{I}),
    \label{eq:edm_forward}
\end{equation}
where $\mathcal{N}(\mu, \Sigma)$ denotes a Gaussian distribution with mean $\mu$ and covariance matrix $\Sigma$, and $\sigma \in [\sigma_{\min}, \sigma_{\max}]$ is drawn from a continuous noise-level distribution. In the limit $\sigma \to \sigma_{\max}$, the perturbed distribution approaches $\mathcal{N}(0, \sigma_{\max}^2 \mathbf{I})$, i.e.\ pure noise.

The central idea of EDM is the introduction of a preconditioned denoiser $D_\theta$, designed to improve training stability by normalising inputs and outputs across noise levels:
\begin{equation}
    D_\theta(\tilde{x};\, \sigma) = c_\mathrm{skip}(\sigma)\,\tilde{x}
    + c_\mathrm{out}(\sigma)\, F_\theta\!\left(c_\mathrm{in}(\sigma)\,\tilde{x};\,
    c_\mathrm{noise}(\sigma)\right),
    \label{eq:edm_precond}
\end{equation}
where $F_\theta$ is the raw neural network and the preconditioning scalars are chosen as:
\begin{equation}
    c_\mathrm{skip}(\sigma) = \frac{\sigma_\mathrm{data}^2}{\sigma^2 + \sigma_\mathrm{data}^2},
    \quad
    c_\mathrm{out}(\sigma) = \frac{\sigma\,\sigma_\mathrm{data}}{\sqrt{\sigma^2 + \sigma_\mathrm{data}^2}},
    \quad
    c_\mathrm{in}(\sigma) = \frac{1}{\sqrt{\sigma^2 + \sigma_\mathrm{data}^2}},
    \quad
    c_\mathrm{noise}(\sigma) = \tfrac{1}{4}\ln \sigma,
\end{equation}
with $\sigma_\mathrm{data}$ denoting the standard deviation of the training data. These scalings ensure that both the input to $F_\theta$ and its output have unit variance at any noise level, which stabilises gradient flow during training.

The training objective minimises a weighted denoising loss over noise levels sampled from a log-normal distribution $\ln \sigma \sim \mathcal{N}(P_\mathrm{mean}, P_\mathrm{std}^2)$:
\begin{equation}
    L(\theta) = \mathbf{E}_{\sigma,\, x_0,\, \mathbf{n} \sim \mathcal{N}(0, \sigma^2\mathbf{I})}
    \!\left[\, \lambda(\sigma)\,
    \left\| D_\theta(x_0 + \mathbf{n};\, \sigma) - x_0 \right\|_2^2 \right],
    \label{eq:edm_loss}
\end{equation}
where $\mathbf{n} \sim \mathcal{N}(0, \sigma^2 \mathbf{I})$ is the Gaussian noise added to the clean sample $x_0$ to produce the noisy input $x_0 + \mathbf{n}$, and the loss weighting $\lambda(\sigma) = (\sigma^2 + \sigma_\mathrm{data}^2)\,/\,(\sigma\,\sigma_\mathrm{data})^2$ up-weights intermediate noise levels, which carry the most learning signal.

At inference time, generation proceeds by integrating either a stochastic differential equation (SDE) or a deterministic probability-flow ordinary differential equation (ODE) from $\sigma_{\max}$ to $\sigma_{\min}$, with the denoiser $D_\theta$ providing the score estimate at each step, $s_\theta(x,\sigma) = (D_\theta(x;\,\sigma) - x)\,/\,\sigma^2$. This continuous-time formulation decouples the number of network evaluations from the training parametrisation, allowing flexible trade-offs between sample quality and computational cost, and its preconditioning coefficients (Eq.~\ref{eq:edm_precond}) ensure that $F_\theta$ sees approximately unit-variance inputs and targets across noise levels.

\subsection{Vision Transformer Architecture}

The Vision Transformer (ViT) \citep{dosovitskiy2020} adapts the Transformer architecture \citep{vaswani2017} to image data by dividing the input into a sequence of non-overlapping patches, linearly projecting each patch into a token embedding, and processing the resulting sequence with standard multi-head self-attention layers. Unlike convolutional networks, which impose a local inductive bias through fixed receptive fields, self-attention operates globally over all patch pairs, allowing the model to capture long-range spatial dependencies from the first layer. Each Transformer block consists of a multi-head self-attention module followed by a position-wise feed-forward network, with layer normalisation and residual connections applied around each sub-layer. Positional information is injected either through learned absolute embeddings or, in more recent variants, through Rotary Position Embeddings (RoPE) \citep{su2024rope}, which encode relative position directly in the attention computation. ViT-based architectures have demonstrated strong performance on a variety of dense prediction tasks and have been increasingly adopted as the backbone $F_\theta$ in diffusion models, where the ability to model global structure is particularly valuable.

\subsection{U-Net Architecture}

The U-Net \citep{ronneberger2015} is an encoder-decoder convolutional architecture originally proposed for biomedical image segmentation, characterised by symmetric downsampling and upsampling paths connected by skip connections. The encoder progressively reduces spatial resolution through strided convolutions or pooling operations, extracting hierarchical feature representations at increasing levels of abstraction. The decoder mirrors this structure, recovering spatial detail through transposed convolutions or interpolation. The defining feature of the U-Net is the direct concatenation of encoder feature maps into the corresponding decoder stages via skip connections, which allows the network to combine high-level semantic information from the bottleneck with fine-grained spatial detail from early layers, avoiding the loss of localisation information that plagues plain encoder-decoder architectures. Due to its ability to model both global context and local structure simultaneously, the U-Net has been widely adopted as the backbone $F_\theta$ in denoising diffusion models \citep{ho2020}, where the task of recovering a clean signal from noise requires reasoning at multiple spatial scales.

\section{Methodology}

\subsection{Problem Formulation}
\label{subsec:problem_formulation}

This study aims to characterise the relationship between solar wind velocity, the three components of the interplanetary magnetic field, and the ionospheric convection pattern, represented by the electric potential maps.
To this end, we train different deep learning models to predict the future state of the ionospheric convection pattern, given the current state, the solar wind velocity and the interplanetary magnetic field vector measured at the Lagrangian point L1, which is located approximately 1.5 million kilometres from Earth in the direction of the Sun.
This formulation treats the models as though positioned at L1 at time $t_0$: the electric potential maps are observed from $t_{-N}$ to $t_0$ ($N$ past frames), while the L1 solar wind/IMF timeseries is observed from $t_{-N}$ to $t_M$ ($M$ frames beyond $t_0$), because the finite L1-to-Earth propagation delay (Section~\ref{subsec:data_alignment}) means the solar wind driver has already been measured at L1 before its effect on the ionosphere becomes observable. The $M$ future electric potential maps, from $t_0$ to $t_M$, are therefore the target of the prediction task.

Selecting an appropriate value for $M$ is crucial, because it defines the time horizon of the prediction task, which is directly related to the L1-to-Earth propagation delay described in Section~\ref{subsec:data_alignment}, approximately 30-60 minutes depending on the solar wind speed.
In our case we select $N$ equals to 15 frames which correspond to 30 minutes, while $M$ equals to 7 frames which correspond to 14 minutes (the resulting 22-frame sequence length was already introduced in Section~\ref{subsec:ml_preprocessing}), since the time resolution of the electric potential maps in our dataset is 2 minutes, as described in Section \ref{subsec:superdarn}.
We refer with the term \textit{trajectory} to the sequence of $M$ future electric potential maps that the models are trained to predict, given the past $N$ electric potential maps and the solar wind and IMF measurements from L1.

\begin{figure}
    \centering
    \includegraphics[width=\linewidth]{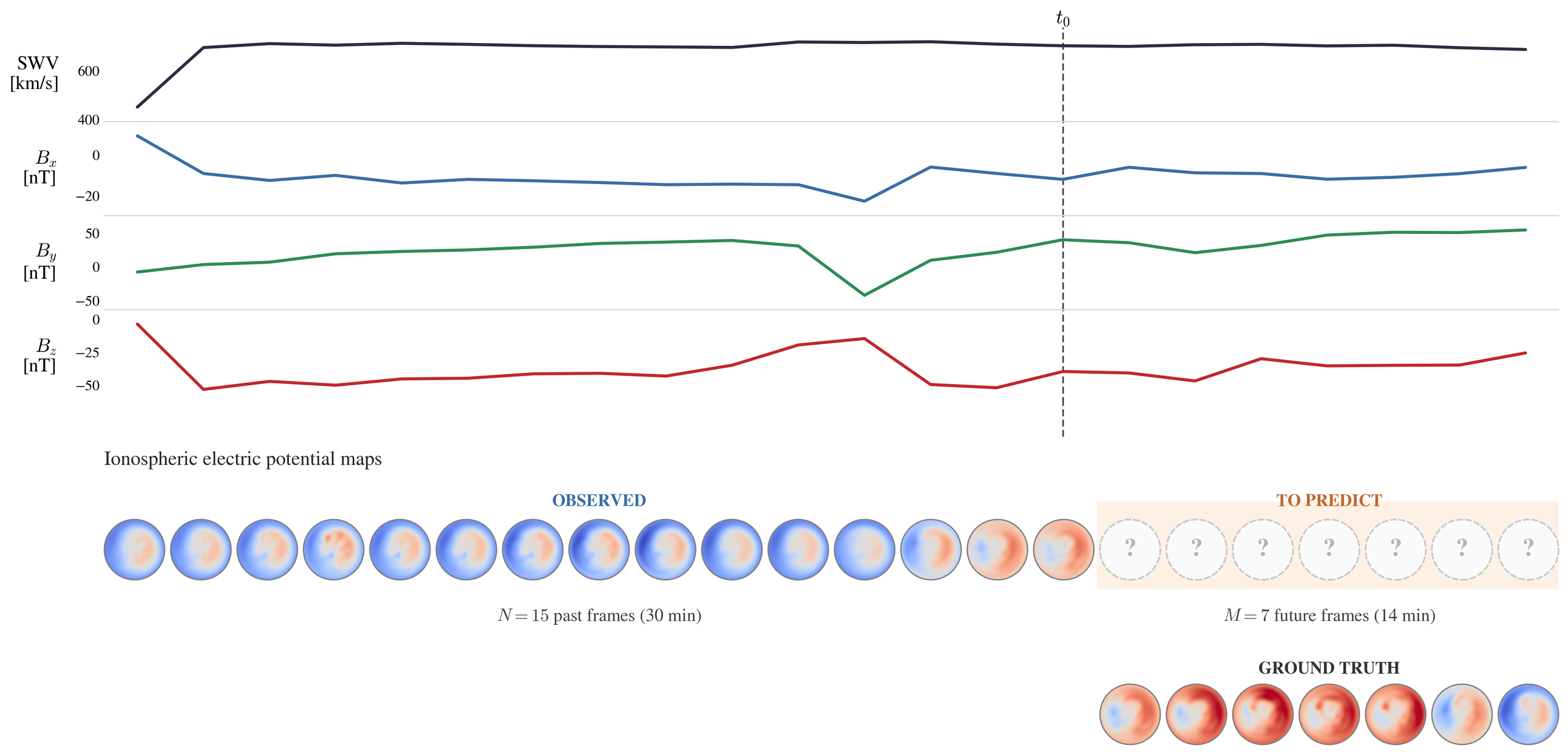}
    \caption{The forecasting task as a multimodal completion problem, illustrated on one real
    training sequence. Top: the four L1 solar wind and IMF timeseries (SWV, $B_x$, $B_y$, $B_z$)
    are observed across the entire window, both past ($t_{-N}$ to $t_0$) and future ($t_0$ to
    $t_M$). Bottom: the corresponding electric potential maps are only known for the past $N=15$
    frames (OBSERVED); the future $M=7$ frames (TO PREDICT, shaded) are the target of the
    prediction task. The bottom row shows the true, unseen ground truth for those $M$ frames for
    reference.}
    \label{fig:training_strategy}
\end{figure}

Figure~\ref{fig:training_strategy} illustrates an example of the training methodology. The four panels at the top show a timeseries of L1 solar wind and IMF timeseries (SWV, $B_x$, $B_y$, $B_z$, top), observed across the full window, both past and future relative to $t_0$, while the electric potential maps (bottom) are only known for the past $N=15$ frames, with the future $M=7$ frames withheld as the prediction target.

This prediction task can also be viewed as a multimodal video forecasting problem (Figure \ref{fig:training_strategy}): unlike conventional video prediction, where future frames are inferred solely from previously observed frames, our model is additionally conditioned on solar wind and interplanetary magnetic field measurements that extend beyond the present time ($t_0$). An intuitive analogy is a movie in which both audio and video are available for the first half of the sequence but only the audio track for the second half, and the objective is to reconstruct the missing video frames from the available soundtrack and prior visual context.
However, unlike standard audio-video prediction, the two modalities are not naturally synchronised: the solar wind measurements reflect the upstream driver before its effects become observable in the ionosphere, so the conditioning signal already contains information about future ionospheric states not yet present in the electric potential maps at ($t_0$). This introduces an additional challenge, absent from conventional video forecasting, in learning the temporal relationship between the two modalities.

\subsection{Training Strategy}
\label{subsec:training_strategy}

We train three models to address the prediction task described above, hereafter referred to as U-Net, CLASSIC, and NOCOND. U-Net is a deterministic U-Net conditioned on the full set of inputs (past electric potential maps and L1 solar wind and IMF measurements).
CLASSIC is a probabilistic diffusion model with a ViT backbone using the same conditioning. NOCOND is a probabilistic diffusion model with a ViT backbone conditioned only on the past electric potential maps,
with no L1 measurements provided.

Although the solar wind and IMF at L1 are well-established as the primary external drivers of ionospheric convection \citep{dungey1961}, the unconditioned model provides a meaningful baseline for assessing the information gained through external conditioning. The ionosphere is not a purely instantaneous response system: it retains its own temporal inertia, meaning that its near-future state is partially determined by its own recent history \citep{cowley1992}. This effect is expected to be most prominent during geomagnetically quiet periods, when solar wind forcing is weak and the convection pattern evolves slowly. By evaluating a model that has access only
to past ionospheric state, and in particular by examining its performance during quiet intervals, we can isolate and quantify the predictive role of ionospheric memory, separately from the contribution of the upstream solar wind driver.

The training and validation sets are constructed as described in Section~\ref{subsec:ml_preprocessing}. All models are trained for 200,000 steps with the AdamW optimizer and a cosine-annealed learning rate schedule: a linear warmup over the first 1,000 steps from $10^{-6}$ to a peak of $10^{-4}$, followed by cosine decay to a minimum of $10^{-7}$ over the remaining steps.
The models are trained on 4 NVIDIA GH200 120GB GPUs, with a batch size of 1 per GPU, for a total of 4 samples per training step, and all models have about 800 million parameters.

The ViT backbone patchifies each $128\times128$ frame with patch size $1\times8\times8$ (time$\times$height$\times$width) into a token sequence processed by 48 transformer blocks of hidden width 1024 with 16 attention heads, QK-normalisation, and rotary position embeddings \citep{su2024rope}. The past electric potential frames are then concatenated as additional input channels alongside the noisy target, while the L1 solar wind and IMF timeseries are mapped through a small MLP and injected via adaptive layer-norm (AdaLN-Zero) modulation at every transformer block, in addition to being broadcast as extra per-pixel channels. EDM preconditioning (Section~\ref{subsec:diffusion}) uses $\sigma_\text{data}=0.5$, with $\sigma_\text{min}=0.01$ and $\sigma_\text{max}=100$ during training. We maintain an exponential moving average (EMA) of the model weights throughout training (inverse power schedule, power 0.667, maximum decay 0.9999), and it is the EMA weights, not the raw training weights, that are used for all inference and evaluation. At inference, CLASSIC and NOCOND samples are generated with the stochastic DPM-Solver++(2M) SDE sampler \citep{lu2022dpmsolverpp} using 50 denoising steps (NFE=50) over a Karras noise schedule ($\sigma_\text{min}=0.01$, $\sigma_\text{max}=80$).

\subsection{Evaluation Under Distinct Geomagnetic Regimes}
\label{subsec:different_geomagnetic_regimes}

\begin{figure}
    \centering
    \includegraphics[width=\linewidth]{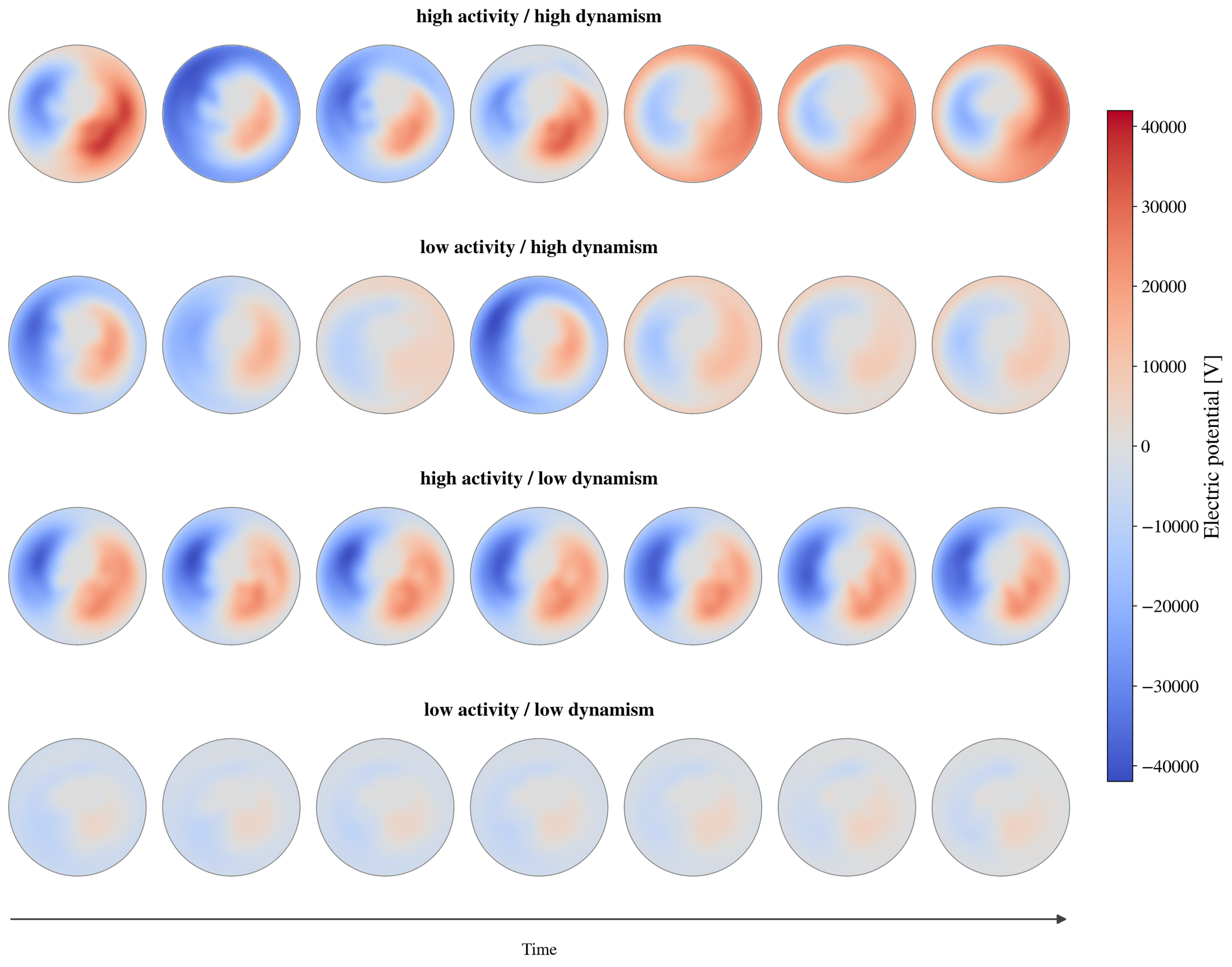}
    \caption{Real example sequences illustrating the four evaluation regimes obtained by
    crossing geomagnetic activity (Akasofu $\epsilon$) with convection dynamism (mean
    frame-to-frame potential change), as defined in this section. For each regime, the most
    extreme validation-set example (largest $|\epsilon|$ and dynamism deviation in the direction
    of that regime's label) is shown as seven evenly spaced snapshots from its input window,
    on a shared colour scale.}
    \label{fig:different_types_trajectories}
\end{figure}

At evaluation time, rather than reporting aggregate metrics over the full validation set, we stratify the results according to two independent dimensions: geomagnetic activity and convection dynamism. This is motivated by the prediction task differing systematically with geomagnetic activity, so a model performing well on average may still fail systematically during particular events.

We quantify geomagnetic \textit{activity} using the Akasofu epsilon parameter \citep{akasofu1981}, a proxy for the electromagnetic energy flux transferred from the solar wind to the magnetosphere, described in Section \ref{subsec:indices}. For each trajectory, we take the maximum epsilon value over the full input window as the representative activity level. A trajectory is classified as \textit{high activity} if this value exceeds the 75th percentile of the validation distribution, and as \textit{low activity} if it falls below the 25th percentile.

We quantify convection dynamism as the mean absolute frame-to-frame difference in the electric potential maps over the frames (in physical units, V/frame). This score measures how rapidly the convection pattern is evolving in the input window, independently of its absolute amplitude. A trajectory is classified as high dynamism if its score falls in the top 25\% of the validation distribution (i.e.\ above the 75th percentile), and as low dynamism if it falls in the bottom 25\% (i.e.\ below the 25th percentile).

Examples trajectories that fall on the different regimes described above are shown in Figure~\ref{fig:different_types_trajectories}. Combining these two binary labels yields four evaluation regimes: high activity / high dynamism (first row from the top), low activity / high dynamism (second row from the top), high activity / low dynamism (third row from the top), and low activity / low dynamism (fourth row from the top). High-dynamism regimes (top two rows) show visibly evolving convection patterns, while low-dynamism regimes (bottom two rows) remain close to fixed in shape; high-activity regimes (rows one and three) show substantially larger potential magnitudes than their low-activity counterparts. This decomposition lets us diagnose model behaviour across distinct physical conditions, e.g.\ whether deterministic and probabilistic approaches are differentially suited to low versus high activity \citep{bishop1994mdn}.

\subsubsection{Single-Pass and Autoregressive Evaluation}
\label{subsubsec:single_pass_vs_autoregressive}

\begin{figure}
    \centering
    \includegraphics[width=\linewidth]{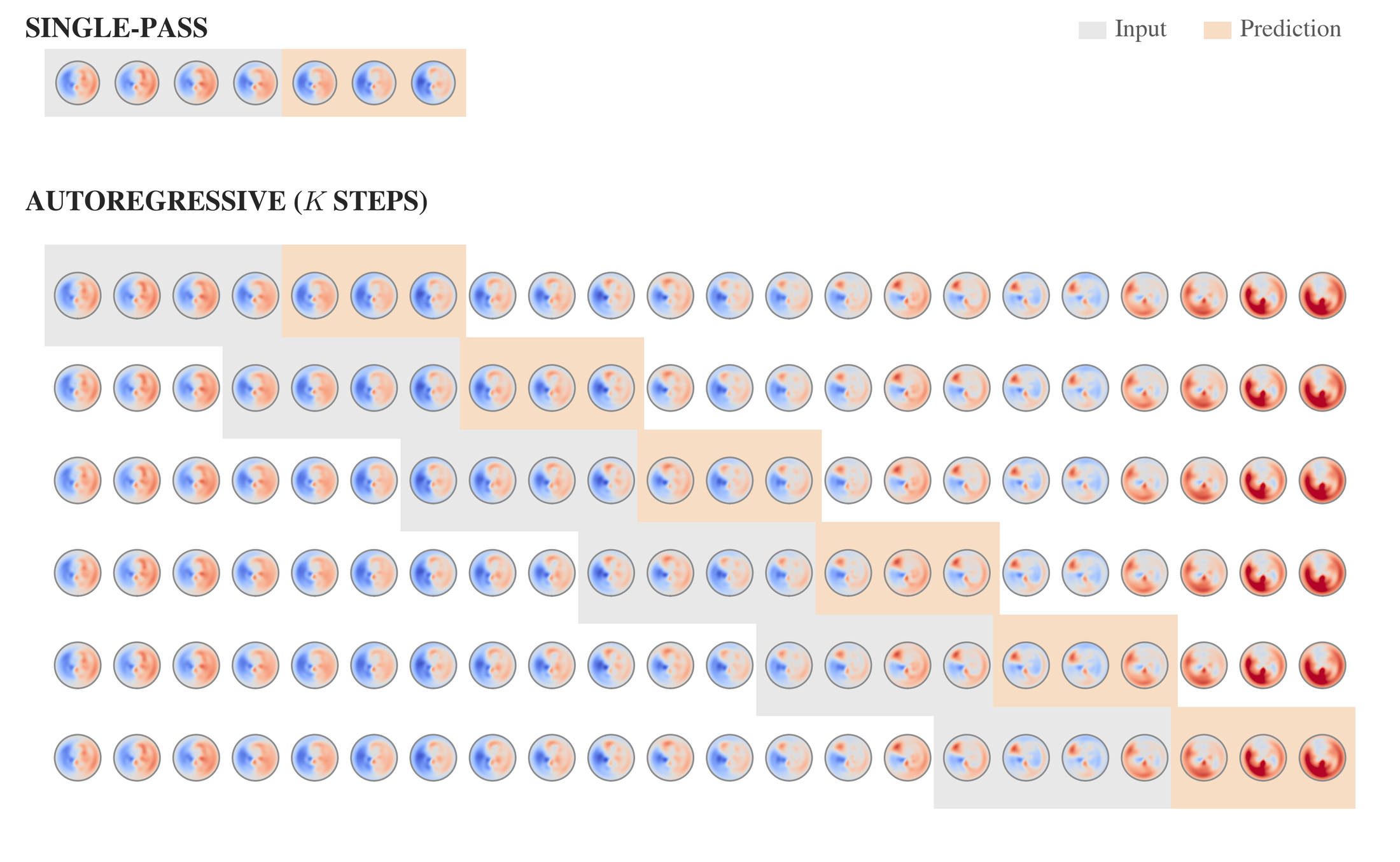}
    \caption{Single-pass versus autoregressive evaluation, illustrated on one real trajectory
    (gray = input frames, orange = predicted frames; window size reduced to 4 input / 3 output
    frames here for legibility, versus the actual $N=15$/$M=7$ used throughout). \textit{Top}: in
    the single-pass regime the model consumes one input window and produces one output window in
    a single forward pass. \textit{Bottom}: in the autoregressive regime the same sliding window is
    applied repeatedly across the full trajectory, each row's predicted frames becoming part of the
    input window for the next row, for a total of $K$ steps.}
    \label{fig:single_step_vs_autoregressive}
\end{figure}

In addition to the activity/dynamism stratification, the evaluation is structured along the prediction horizon. The schematics is illustrated in Figure~\ref{fig:single_step_vs_autoregressive}.

The first is the single-pass regime (Figure~\ref{fig:single_step_vs_autoregressive}, top panel), which visualises the training. The model receives $N$ past electric potential maps and the corresponding L1 timeseries and predicts the next $M$ frames in a single forward pass. This regime provides the most direct assessment of what the model has learned, as the task at inference time is identical to the one seen during training.

The second is the autoregressive regime (Figure~\ref{fig:single_step_vs_autoregressive}, bottom panel), designed to test the stability and generalisation of the model beyond its training objective. At each step, the $M$ predicted frames are appended to the input window, which is then shifted forward by $M$ frames to form the context for the next prediction, repeating this process for a total of $K$ steps and producing a trajectory of $K \cdot M$ future frames. This rollout is autoregressive only with respect to the electric potential maps: the L1 solar wind and IMF timeseries conditioning every step are always real, observed data, never predicted. Since the model was never trained to condition on its own predictions, this regime constitutes an out-of-distribution stress test with respect to the task: any error accumulated in the predicted maps is fed back as input, revealing whether the model remains stable over extended horizons or whether its outputs progressively degrade.

\subsection{Out-of-Distribution Case Study: St. Patrick's Day Storm (March 2015)}
\label{subsec:st_patrick_test}

\begin{figure}
    \centering
    \includegraphics[width=0.6\linewidth]{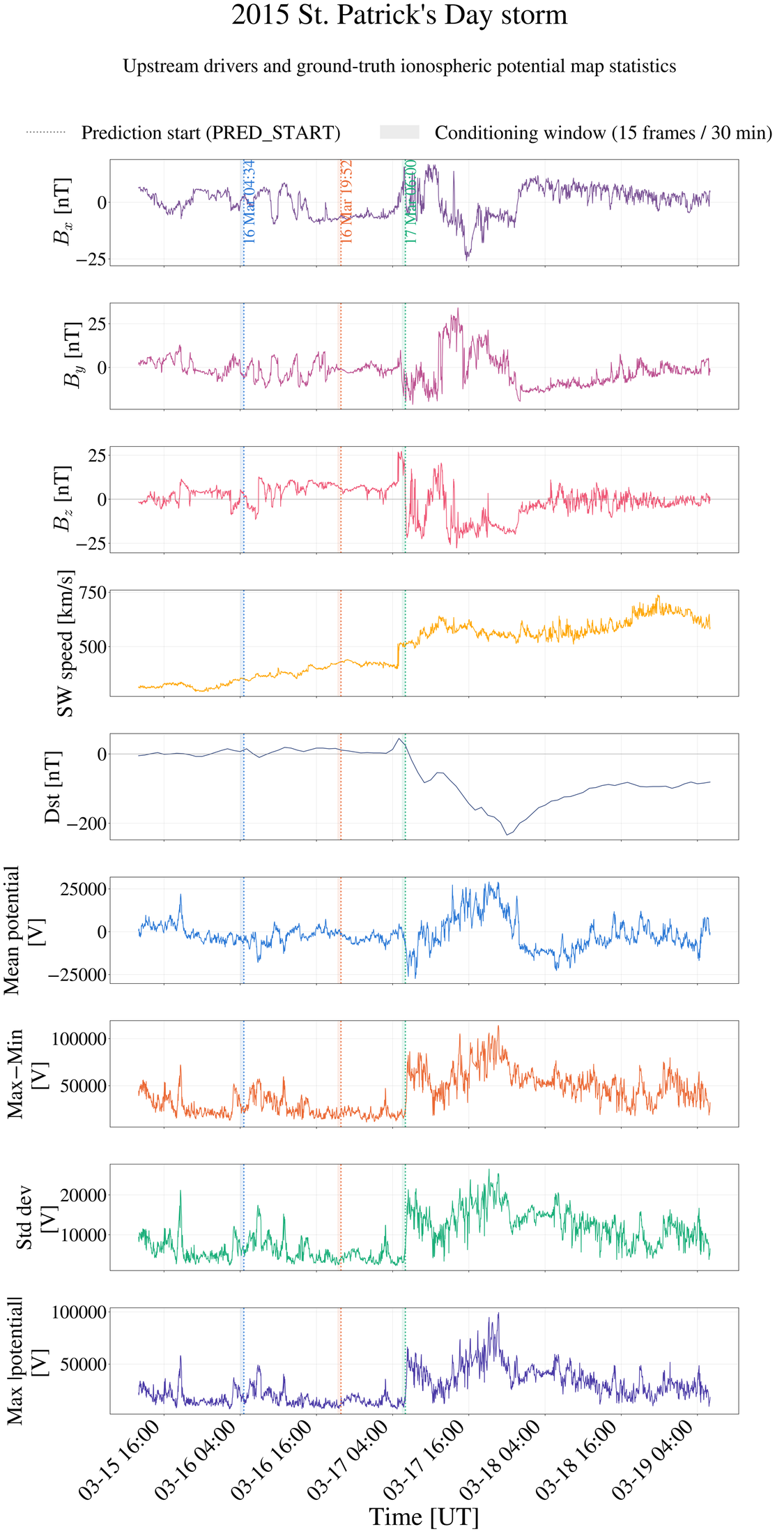}
    \caption{Upstream drivers (IMF $B_x$, $B_y$, $B_z$; solar wind speed; Dst index) and observed
    ionospheric potential map statistics (spatial mean, max$-$min, standard deviation, max$|\Phi|$)
    over the St.\ Patrick's Day storm, 15--19 March 2015. Vertical dotted lines and shaded bands mark
    the three \texttt{PRED\_START} windows used for autoregressive generation
    (Section~\ref{subsec:results_2015}) and their 30-minute conditioning window. The third window (17
    March 06:00) falls on the main-phase onset: $B_z$ turns sharply southward, solar wind speed jumps,
    Dst begins its plunge toward its storm minimum, and all four ionospheric statistics show a sharp
    jump immediately after the marker; the other two windows sample the pre-storm quiet interval.}
    \label{fig:st_patrick_event}
\end{figure}

As a final evaluation, we analyse the St.\ Patrick's Day storm of 17 March 2015, a specific geomagnetic event that lies entirely outside the training and validation periods. Figure~\ref{fig:st_patrick_event} places this event in context, showing the upstream IMF components, solar wind speed, and Dst index alongside the observed ionospheric potential map statistics across the storm period, together with the three windows used to initialise the autoregressive rollouts evaluated in Section~\ref{subsec:results_2015}.
This event is selected for two compounding reasons that make it a stringent out-of-distribution test case.

First, the storm belongs to Solar Cycle 24, whereas all training and validation data are drawn from Solar Cycle 25 (2020--2025). Consecutive solar cycles differ in their overall intensity and in the frequency and magnitude of energetic events, so evaluating the models on an event from a different cycle tests whether the learned relationship between upstream solar wind forcing and ionospheric response generalises beyond the specific activity regime of the training period. Second, the St.\ Patrick's Day storm is one of the most intense geomagnetic storms of Solar Cycle 24, with a minimum Dst of approximately $-234$ nT, classifying it as an intense storm by the \citet{gonzalez1994} scale (Dst $\leq -100$~nT). The associated ionospheric convection pattern was highly disturbed and rapidly evolving, placing this event in the extreme tail of both the activity and dynamism regimes defined in Section \ref{subsec:different_geomagnetic_regimes}. The high-latitude ionospheric response to this specific storm is independently well documented, from GNSS scintillation and irregularity observations in both hemispheres \citep{dangelo2018} to storm-phase-resolved studies of the high-latitude convection electric field derived from SuperDARN \citep{Astafyeva2015}, providing external points of reference against which the qualitative behaviour of our models in Section~\ref{subsec:results_2015} can be compared.

By applying the models to this event without any fine-tuning or adaptation, we assess whether the learned representations generalise beyond their training cycle and whether the probabilistic models, in particular, retain calibrated uncertainty estimates under conditions of extreme and unfamiliar forcing. This case study therefore serves both as a robustness probe and as a qualitative illustration of model behaviour at the boundary of what the training distribution can support.

\subsection{Evaluation Metrics}
\label{subsec:metrics}

We evaluate model performance using two complementary families of metrics. To capture global dynamic range, we use the normalised errors on scalar summary statistics of the electric potential field and to capture spatial structure we use image-quality metrics computed across the predicted maps. For the probabilistic diffusion models, which generate $S$ samples per trajectory ($S=10$ for the single-pass and $K=50$ autoregressive evaluations, $S=3$ for the St.\ Patrick's Day storm case study, reduced due to the computational cost of the longer multi-day rollout), every metric below is first computed independently for each sample and then summarised as a mean $\pm$ one standard deviation across samples. This isolates the sample-to-sample (epistemic) spread of the model from the frame-to-frame variability already absorbed by the pooling operation used within each metric. The deterministic U-Net and the persistence baseline (Section~\ref{subsec:metrics}) yield a single sample and are reported without an associated spread.

\subsubsection{Normalised Root Mean Square Error of Summary Statistics}
\label{subsubsec:nrmse}

For each electric potential map $\Phi$ we define three scalar summary statistics: the spatial mean $\bar{\Phi} = \langle \Phi(\mathbf{x}) \rangle_{\mathbf{x}}$, the spatial standard deviation $\sigma_{\Phi}$, and the Interquartile Potential Range (IPR, Section~\ref{subsubsec:ipr}). For each statistic $X \in \{\bar{\Phi}, \sigma_{\Phi}, \text{IPR}\}$ we track its evolution across the evaluated frames of a trajectory (or, in the single-pass regime, across the evaluation set) and report the normalised root mean square error between the predicted and observed (from now on "ground-truth") time series,
\begin{equation}
  \text{NRMSE}_{X} = \frac{\sqrt{\left\langle \left(X^{\text{pred}} - X^{\text{gt}}\right)^2 \right\rangle}}{\sigma\!\left(X^{\text{gt}}\right)},
  \label{eq:nrmse}
\end{equation}
where $\langle \cdot \rangle$ denotes the average over all evaluated frames pooled within an evaluation group (e.g.\ a given geomagnetic regime, or a given prediction window in the St.\ Patrick's Day case study) and $\sigma(X^{\text{gt}})$ is the standard deviation of the ground-truth values of $X$ over that same pooled set. Normalising by $\sigma(X^{\text{gt}})$ makes $\text{NRMSE}_X$ comparable across statistics with different physical scales and expresses the error in units of the ground truth's natural variability: values near zero indicate the model tracks the true evolution of $X$, while $\text{NRMSE}_X \gtrsim 1$ indicates an error comparable to or larger than that variability. As a reference, we additionally report $\text{NRMSE}_X$ for a persistence baseline, obtained by repeating the last conditioning frame for each of the $M$ predicted frames. We expect a competitive model to outperform this baseline, especially at longer lead times.

\subsubsection{Interquartile Potential Range (IPR)}
\label{subsubsec:ipr}

Convection intensity is conventionally quantified via the Cross-Polar Cap Potential (CPCP), defined as the difference between the maximum and minimum of the electric potential map across the polar cap \citep{boyle1997}, but this two-pixel-extrema definition is highly sensitive to localised noise, data fitting methods or artifacts in predicted maps. We therefore adopt a robust variant, the Interquartile Potential Range,
\begin{equation}
  \text{IPR} = \Phi_{P75} - \Phi_{P25},
\end{equation}
where $\Phi_{P75}$ and $\Phi_{P25}$ are the 75th and 25th percentiles of the potential field, respectively, retaining CPCP's physical interpretation while discarding the extreme tails that make it outlier-sensitive.

\subsubsection{Structural Similarity Index (SSIM)}
\label{subsubsec:ssim}

SSIM \citep{wang2004ssim} compares local luminance, contrast, and structure between the predicted and ground-truth maps over a sliding window,
\begin{equation}
  \text{SSIM}(\Phi^{\text{pred}}, \Phi^{\text{gt}}) =
  \frac{(2\mu_{\text{pred}}\mu_{\text{gt}} + c_1)(2\sigma_{\text{pred,gt}} + c_2)}
       {(\mu_{\text{pred}}^2 + \mu_{\text{gt}}^2 + c_1)(\sigma_{\text{pred}}^2 + \sigma_{\text{gt}}^2 + c_2)},
\end{equation}
where $\mu$ and $\sigma$ denote local means and standard deviations, $\sigma_{\text{pred,gt}}$ the local covariance, and $c_1, c_2$ constants for numerical stability. SSIM $\in [-1, 1]$, with $1$ indicating a perfect structural match. Unlike the scalar statistics of Section~\ref{subsubsec:nrmse}, SSIM is sensitive to the spatial arrangement of the convection pattern, not only its global intensity.

\subsubsection{Learned Perceptual Image Patch Similarity (LPIPS)}
\label{subsubsec:lpips}

LPIPS \citep{zhang2018lpips} measures perceptual similarity as a distance in the feature space of a pretrained convolutional network (we use AlexNet), rather than in pixel space,
\begin{equation}
  \text{LPIPS}(\Phi^{\text{pred}}, \Phi^{\text{gt}}) =
  \sum_{l} \frac{1}{H_l W_l} \sum_{h,w} \left\lVert w_l \odot \left(\hat{f}^{l,\text{pred}}_{hw} - \hat{f}^{l,\text{gt}}_{hw}\right) \right\rVert_2^2,
\end{equation}
where $\hat{f}^{l}$ are unit-normalised feature activations at layer $l$, $H_l, W_l$ the spatial dimensions of that layer, and $w_l$ a learned per-channel weighting. Because it operates on deep features rather than raw pixels, LPIPS penalises perceptually salient structural discrepancies (e.g.\ blurring, checkerboard artifacts, missing fine structure) that pixel-wise metrics such as SSIM can under-penalise.

\subsubsection{Power Spectrum Log-MSE (PS log-MSE)}
\label{subsubsec:logmseps}

To assess whether the predicted maps reproduce the correct spatial frequency content, we compute the isotropic power spectrum of each predicted and ground-truth map: each map is Tukey-windowed, Fourier-transformed, and the squared Fourier amplitudes are azimuthally averaged into radial spatial-frequency bins, yielding a 1D spectrum $P^{\text{pred}}_i$ and $P^{\text{gt}}_i$ for the $i$-th frequency bin of the predicted and ground-truth map respectively. We then measure their discrepancy in log-space,
\begin{equation}
  \text{PS log-MSE} = \frac{1}{N}\sum_{i=1}^{N}
  \left(\log_{10}(P^{\text{pred}}_i + \varepsilon) - \log_{10}(P^{\text{gt}}_i + \varepsilon)\right)^2,
\end{equation}
where $N$ is the total number of frequency bins in the power spectrum and $\varepsilon = 10^{-10}$ is a constant added for numerical stability.
Working in log-space gives equal weight to all scales, preventing the dominant low-frequency energy from masking errors at smaller spatial scales. This metric is particularly relevant for detecting the over-smoothing tendency of deterministic models, which tend to suppress high-frequency structures.

\section{Results and Discussion}
\label{sec:results}

We organise the evaluation into three complementary tests that progressively relax the conditions of model training. Section~\ref{subsec:results_singlepass} evaluates the single-pass setting used during training itself, isolating each model's raw predictive skill from any effect of error accumulation. Section~\ref{subsec:results_fakear} extends this to an 11h40 autoregressive rollout, probing whether errors compound once the models must condition on their own output. Section~\ref{subsec:results_2015} pushes this further still, to a multi-day autoregressive rollout on an out-of-distribution storm. Together the three tests trace how the balance between the deterministic U-Net and the probabilistic diffusion models shifts as the prediction horizon lengthens and error is allowed to accumulate. 


\subsection{Single-Pass Evaluation}
\label{subsec:results_singlepass}

In the single-pass regime the models are evaluated using 15 context frames and 7 predicted frames (14 minutes at 2-minute resolution), with no autoregressive feedback and therefore no possibility of error accumulation by construction, which is identical to the training scheme. Results are stratified across the four activity$\times$dynamism regimes defined in Section~\ref{subsec:different_geomagnetic_regimes}, with metrics computed and reported as described in Section~\ref{subsec:metrics}. Unlike the autoregressive case study of Section~\ref{subsec:results_2015}, we do not report the persistence baseline here becauseit is close to trivially competitive (Section~\ref{subsec:results_singlepass_lowdyn}) over a single 14-minute window and it would not add information beyond what the activity/dynamism stratification already shows.

\begin{figure}
    \centering
    \includegraphics[width=\linewidth]{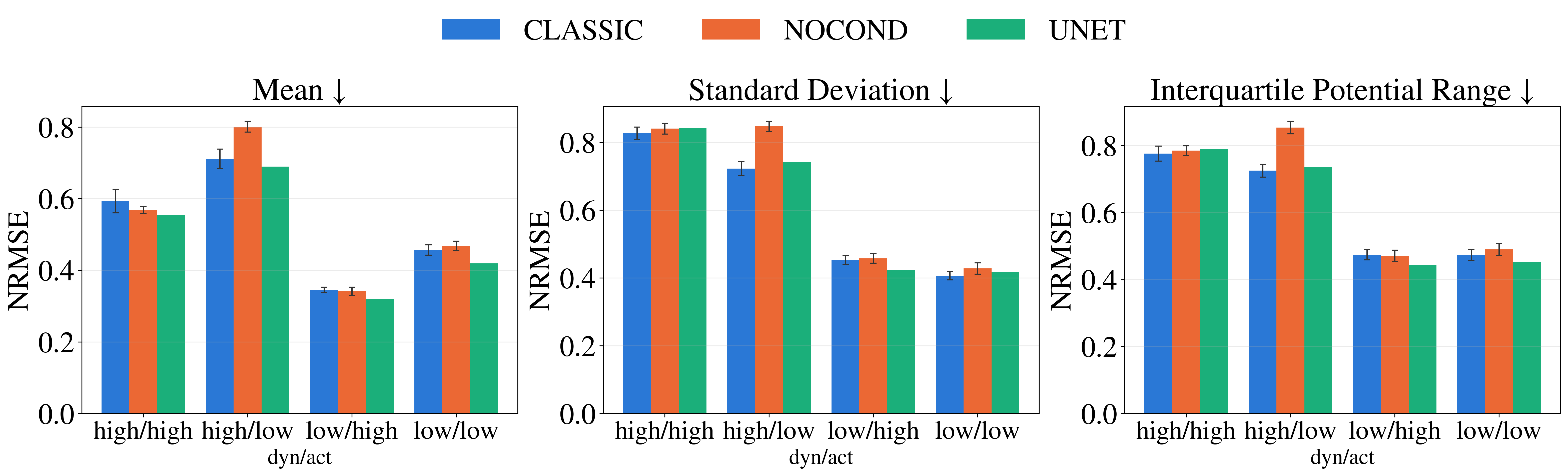}
    \caption{NRMSE of the mean potential, spatial standard deviation, and Interquartile Potential Range
    (IPR), averaged over the full 7-frame prediction window, for each of the four
    activity$\times$dynamism regimes. Here and in subsequent bar-chart figures, $\downarrow$/$\uparrow$
    next to a panel title indicate whether lower or higher values of that metric are better.}
    \label{fig:singlepass_nrmse_overall}
\end{figure}

\begin{figure}
    \centering
    \includegraphics[width=\linewidth]{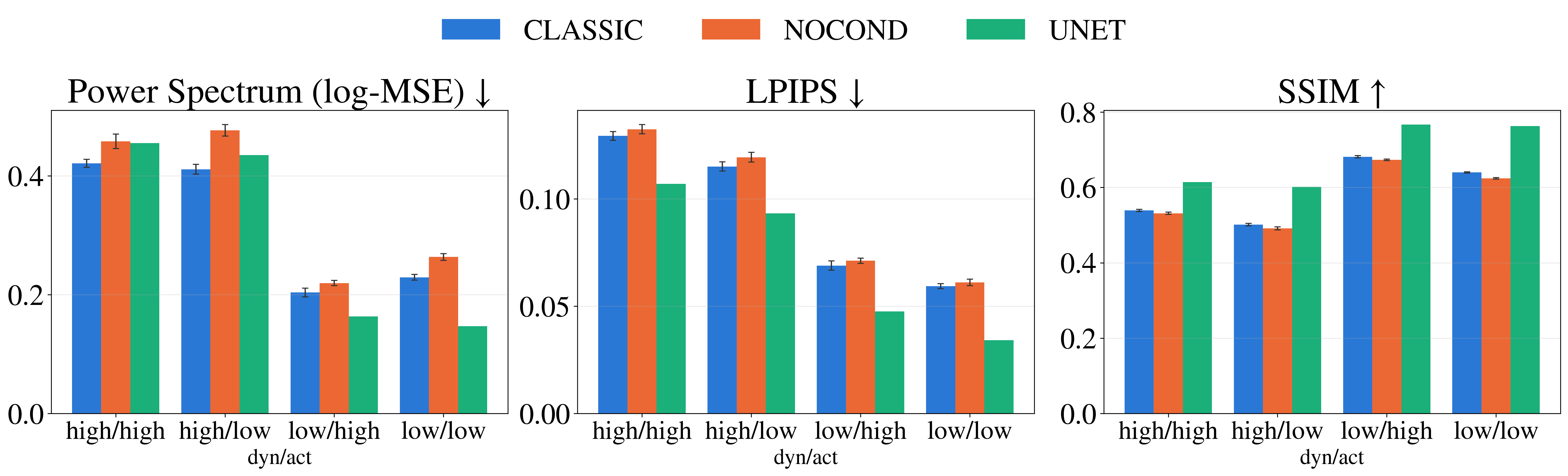}
    \caption{PS log-MSE, LPIPS, and SSIM, averaged over the full 7-frame prediction window, for each of
    the four activity$\times$dynamism regimes.}
    \label{fig:singlepass_bar_overall}
\end{figure}

\begin{figure}
    \centering
    \includegraphics[width=\linewidth]{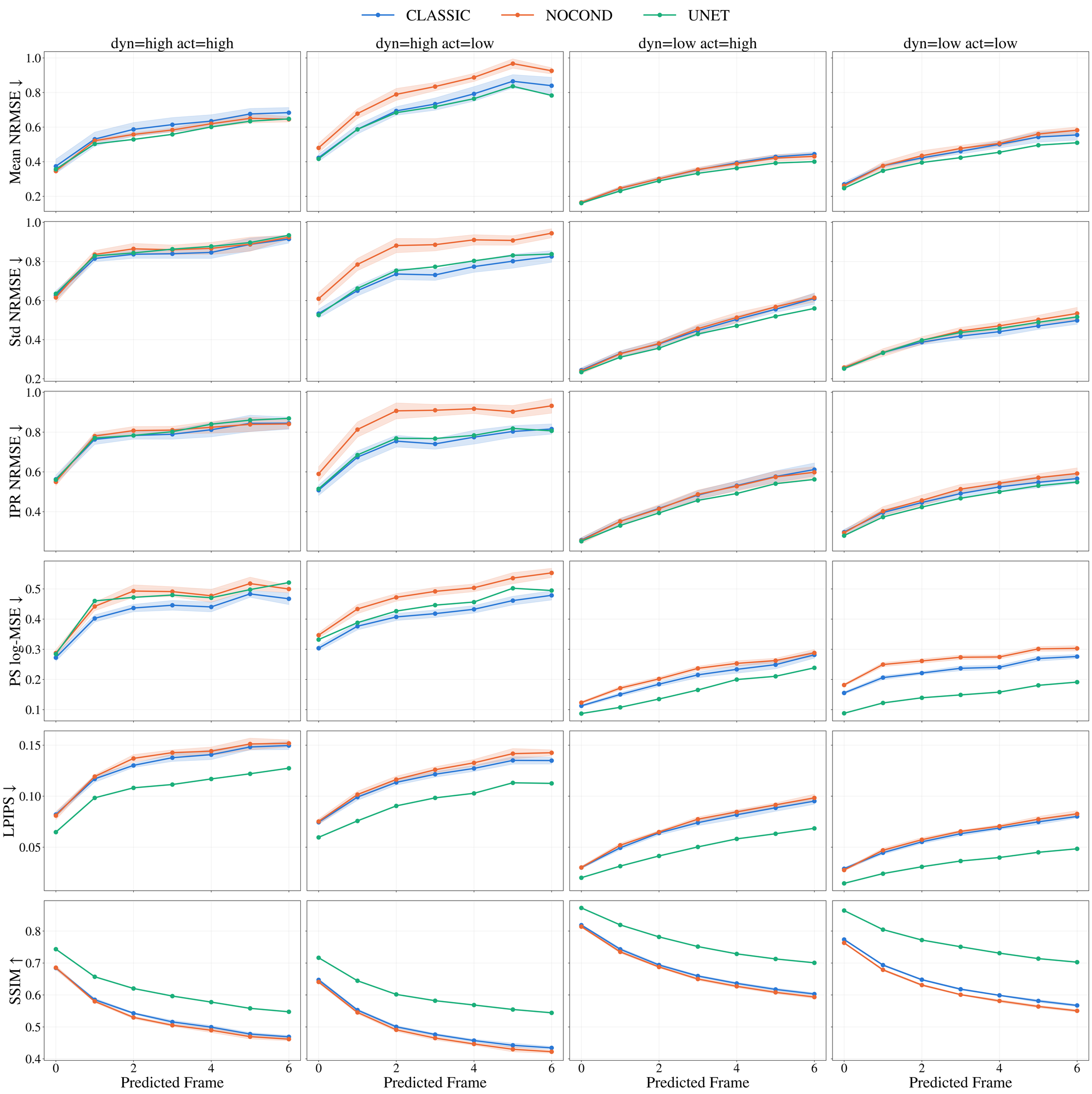}
    \caption{Evolution of NRMSE(mean/std/IPR) and PS log-MSE/LPIPS/SSIM across the 7 predicted frames
    (14 minutes), for each of the four activity$\times$dynamism regimes (columns). Shaded bands denote
    $\pm 1$ standard deviation across diffusion samples for CLASSIC/NOCOND where available.}
    \label{fig:singlepass_degradation}
\end{figure}

\textbf{Global scalar statistics.}
Across all four regimes, NRMSE is broadly similar between the three models with the notable exception of high activity / low dynamics conditions where NOCOND is performing worse than the other two (e.g. Figure~\ref{fig:singlepass_nrmse_overall};
see also the top three rows of Figure~\ref{fig:singlepass_degradation}), differences are on the order of a few percent and, while not perfectly overlapping in every regime, do not consistently favour any one architecture. We attribute this to the short prediction horizon relative to the characteristic timescale of ionospheric convection dynamics. Substorm onset and large-scale
reconfiguration of the convection pattern typically unfold over tens of minutes to hours, so even a sequence classified as high-activity/high-dynamism by our stratification only captures the onset of a change within its 14-minute window, but not its full development. Under these conditions the global amplitude of the field is comparatively easy to predict for every model, and the bulk scalar
statistics carry little discriminative power.

\textbf{Structural and spectral metrics.} SSIM and LPIPS consistently favour U-Net across all four regimes, while PS log-MSE favours CLASSIC in the two more dynamic/active regimes (Figure~\ref{fig:singlepass_bar_overall}; bottom three rows of Figure~\ref{fig:singlepass_degradation}). This pattern reverses in the calmest regime, as discussed below (Section~\ref{subsec:results_singlepass_lowdyn}). We interpret this divergence through the different sensitivities of the two metric families. SSIM/LPIPS are computed against a single predicted realisation and are sensitive to exact spatial alignment. A deterministic model trained to minimise pixel-wise error is expected to regress toward the conditional mean, which is, by definition, the point estimate that minimises expected squared error against any single ground-truth realisation. This systematically favours U-Net when compared against one stochastic sample from a diffusion model, which needs not, and generally does not, coincide pixel-for-pixel with the observed outcome. PS log-MSE, by contrast, is blind to spatial phase as it measures only the amount of power present at each spatial frequency, not its location. This rewards the realistic high-frequency texture retained by individual diffusion samples and penalises the over-smoothing that the regression-to-the-mean behaviour of U-Net induces, even where the diffusion sample is not pixel-aligned with the ground truth. We stress that comparing a deterministic point estimate against individual stochastic samples via SSIM/LPIPS is not a like-for-like comparison, and this limitation should be kept in mind when interpreting Figure~\ref{fig:singlepass_bar_overall}.

\textbf{Dynamism dependence.} \label{subsec:results_singlepass_lowdyn} Regime membership (Section~\ref{subsec:different_geomagnetic_regimes}) is a coarse, binary label and within any one regime, individual sequences still vary considerably in how much the field changes over their 7 predicted frames. In the low-activity/low-dynamism regime, U-Net is the strongest model on every metric reported (Figures~\ref{fig:singlepass_nrmse_overall}--\ref{fig:singlepass_degradation}, right-most column). We interpret this as the same regression-to-the-mean behaviour discussed above for SSIM/LPIPS, but now acting on the model's own temporal inertia rather than on a single spatial realisation. Because U-Net's prediction is biased toward a smoothed extrapolation of the recent state, its frame-to-frame variation is systematically smaller than that of any individual diffusion sample. When the convection pattern does not change much within the 14-minute window, as is, by definition, typical of the low-activity/low-dynamism regime, this smoothing bias is not detrimental but an accurate prior. Predicting little change is, on average, the correct prediction, whereas each diffusion sample still introduces sampling variability that the true (near-static) field does not warrant. CLASSIC's stochastic diversity is therefore advantageous  where the field is uncertain and disadvantageous where it is not.

\begin{figure}
    \centering
    \includegraphics[width=\linewidth]{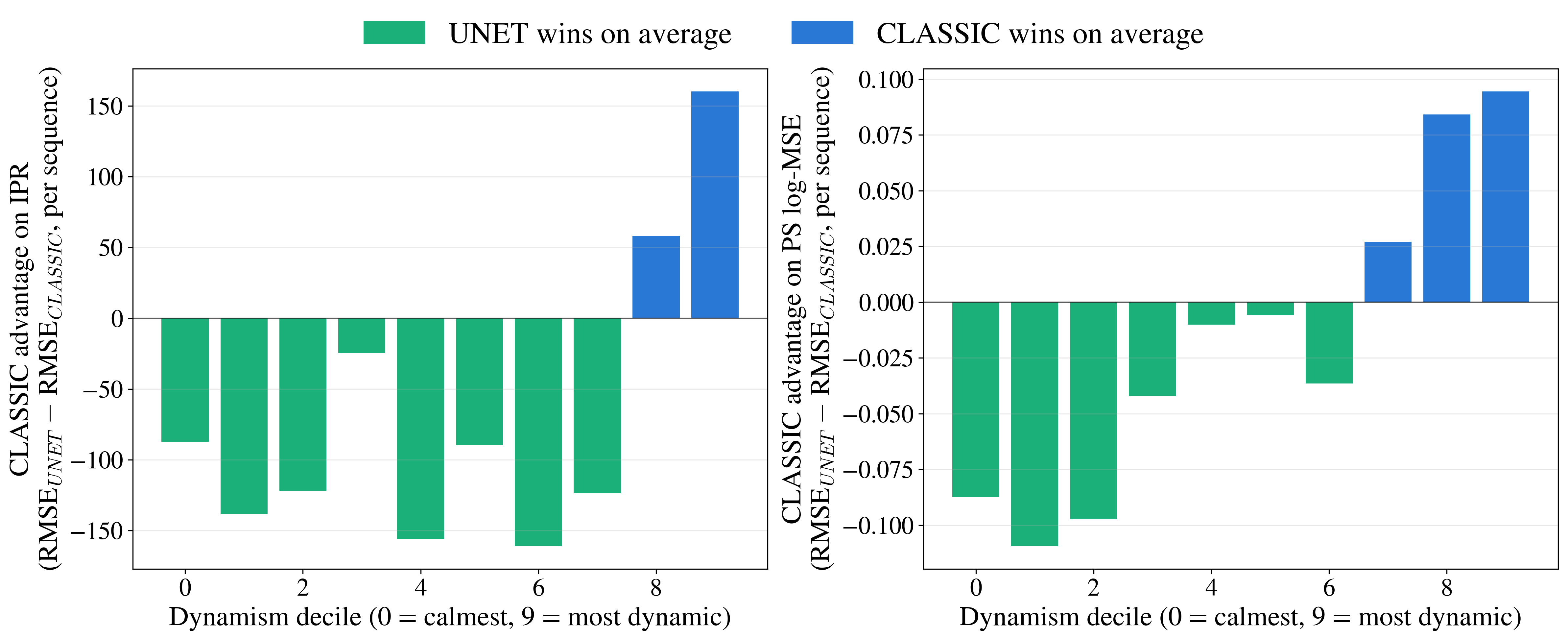}
    \caption{CLASSIC's advantage over U-Net, per sequence, on IPR and PS log-MSE
    (RMSE$_{\text{U-Net}}$ $-$ RMSE$_{\text{CLASSIC}}$, each computed by comparing that model's prediction
    to ground truth over the 7 predicted frames of the sequence; positive favours CLASSIC), binned by
    decile of true per-sequence dynamism and pooled across all four activity$\times$dynamism regimes
    ($n=2260$ sequences: 477 each for the two high-activity regimes, 653 each for the two low-activity
    regimes). Bars are coloured by which model wins on average in that decile
    (green = U-Net, blue = CLASSIC, consistent with all other figures in this section).}
    \label{fig:singlepass_dynamism_decile_all}
\end{figure}

To test this beyond the two regimes discussed above, we define, for each sequence $i$, a per-sequence
dynamism score from its ground-truth spatial mean $\bar{\Phi}_i(t)$ over the 7 predicted frames,
\begin{equation}
  D_i = \max_{t=1,\dots,7} \bar{\Phi}_i(t) - \min_{t=1,\dots,7} \bar{\Phi}_i(t),
\end{equation}
and, for each scalar quantity $X \in \{\text{IPR}, \text{PS log-MSE}\}$, a per-sequence advantage of
CLASSIC over U-Net,
\begin{equation}
  A_i^X = \text{RMSE}_i^{X,\text{U-Net}} - \text{RMSE}_i^{X,\text{CLASSIC}}, \qquad
  \text{RMSE}_i^{X,m} = \sqrt{\frac{1}{7}\sum_{t=1}^{7}\left(X^{m}_{\text{pred}}(i,t) - X_{\text{gt}}(i,t)\right)^2},
\end{equation}
where $\text{RMSE}_i^{X,m}$ compares model $m$'s prediction of $X$ against ground truth over the 7
predicted frames of sequence $i$ alone (for CLASSIC, $X^{m}_{\text{pred}}$ is first averaged over the
$S$ generated samples at each frame, consistent with Section~\ref{subsubsec:nrmse}). By construction,
$A_i^X > 0$ whenever U-Net's error on sequence $i$ exceeds CLASSIC's, i.e.\ CLASSIC is the more accurate
model on that specific sequence; $A_i^X < 0$ indicates the opposite, U-Net more accurate; $A_i^X
\approx 0$ indicates the two models perform comparably on that sequence. Figure~\ref{fig:singlepass_dynamism_decile_all} bins $A_i^X$ by decile of $D_i$, pooling sequences from all four
activity$\times$dynamism regimes.

The sign of the average advantage flips from favouring U-Net in the calmest deciles to favouring CLASSIC in the most dynamic ones.
This result confirms that the pattern is not specific to the two regimes discussed above but is as a continuous trend across the full range of dynamism probed by the validation set and that the appropriate choice is regime-dependent.

These results indicate that the single-pass regime is structurally limited in its ability to expose the differences between deterministic and probabilistic approaches. The prediction horizon is too short, for any meaningful error accumulation to occur, and the dominant factor separating the models is not architecture in the abstract but how well each one's inductive bias (smooth extrapolation for U-Net, stochastic diversity for CLASSIC/NOCOND) matches the true dynamism of the specific 14-minute window being predicted. This motivates an extended-horizon autoregressive rollout test of Section~\ref{subsec:results_fakear}, where the same models are forced to condition on their own predictions over a much longer horizon.

\subsection{Extended-Horizon Autoregressive Rollout}
\label{subsec:results_fakear}

The single-pass evaluation of Section~\ref{subsec:results_singlepass} showed that neither architecture is
uniformly preferable. It is unclear whether that balance is stable as the prediction horizon grows and a 14-minute window is too short for autoregressive error accumulation to play a role. We
therefore repeat the same evaluation under a much longer autoregressive rollout, to test directly whether,
and how, the balance between the deterministic and probabilistic models shifts once errors are allowed to
compound over many steps rather than one.

As in Section~\ref{subsubsec:single_pass_vs_autoregressive}, only the potential map is fed back
autoregressively; the L1 driver remains observed ground truth throughout.
We run $K=50$ AR steps, 350 predicted frames, 11h40, stratified across the same four
activity$\times$dynamism regimes and with the same metric suite and per-sample reporting convention
as the single-pass evaluation (Section~\ref{subsec:results_singlepass}). Unlike the single-pass windows,
a $K=50$ rollout requires 350 consecutive frames with no missing data anywhere in the window, a much
stricter continuity requirement over an 11h40 span than over the 14-minute single-pass window. We
restrict the analysis to the sequences which match this requirement and completed by all three models within
each regime, which is why the number of evaluated sequences is substantially smaller than at single-pass.

\begin{figure}
    \centering
    \includegraphics[width=\linewidth]{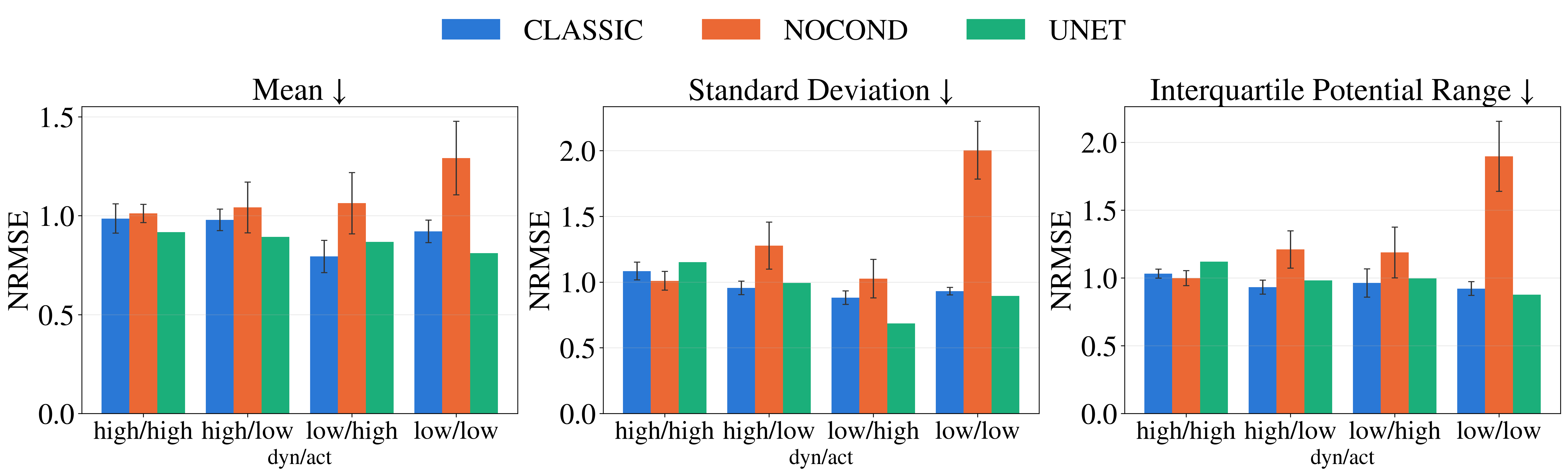}
    \caption{NRMSE of the mean potential, spatial standard deviation, and IPR, averaged over the full
    350-frame ($K=50$, 11h40) rollout, for each of the four activity$\times$dynamism regimes.}
    \label{fig:ark50_nrmse_overall}
\end{figure}

\begin{figure}
    \centering
    \includegraphics[width=\linewidth]{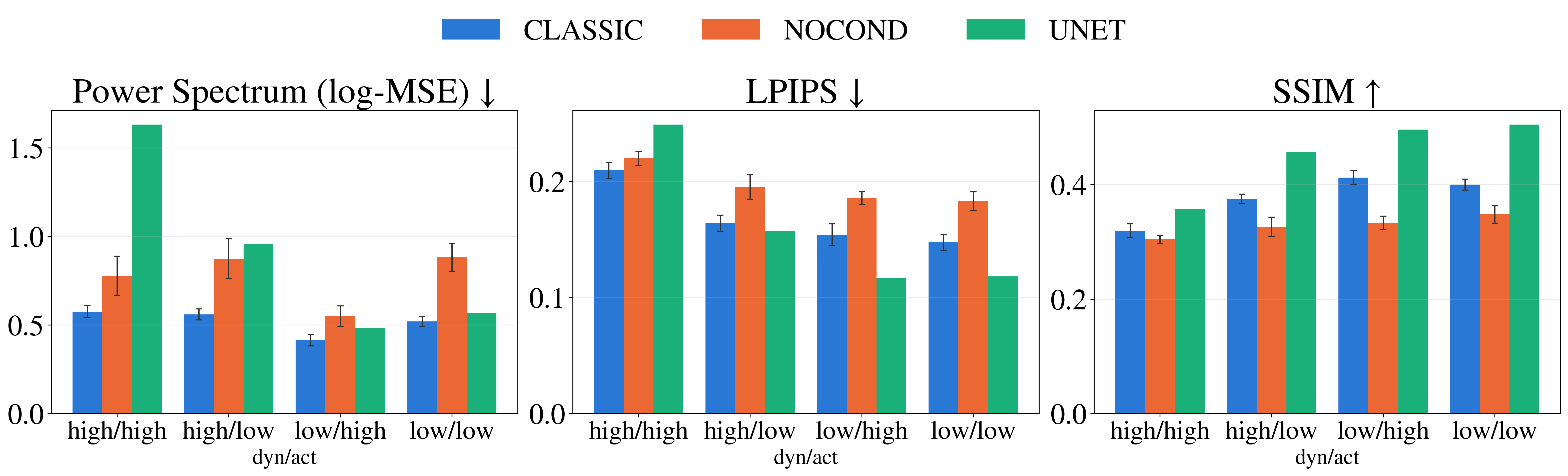}
    \caption{PS log-MSE, LPIPS, and SSIM, averaged over the full 350-frame ($K=50$, 11h40) rollout, for
    each of the four activity$\times$dynamism regimes.}
    \label{fig:ark50_bar_overall}
\end{figure}

\textbf{Global scalar statistics.} Pooling across the
four regimes, U-Net retains a small lead on NRMSE(mean) and NRMSE(std), while the two models are
effectively tied on NRMSE(IPR), CLASSIC marginally ahead (Figure~\ref{fig:ark50_nrmse_overall}). Feeding
predictions back as input for 11h40 makes the scalar-statistic task substantially harder for every model.
NRMSE(mean) increases markedly relative to single-pass, but U-Net still leads, as at single-pass, though now by a smaller margin.

\textbf{Spectral fidelity.} The single-pass evaluation
found PS log-MSE favouring CLASSIC in only three of the four regimes, with U-Net still ahead overall (0.299
vs.\ 0.317, i.e.\ CLASSIC 6.1\% worse). At $K=50$ this changes: CLASSIC's pooled PS log-MSE is 0.518 against U-Net's 0.910, i.e.\ U-Net is 76\% worse (Figure~\ref{fig:ark50_bar_overall}).
This is consistent with the mechanism anticipated at the end of Section~\ref{subsec:results_singlepass}.
Once the model must condition on its own output over many steps, U-Net's regression-to-the-mean bias is no longer a one-shot smoothing of a single 14-minute window but is fed back into every subsequent step, compounding across the full 11h40 rollout, while each CLASSIC sample keeps regenerating plausible high-frequency texture at every AR step rather than inheriting an increasingly smoothed state from itself.

\textbf{Structural similarity.} Pooled across regimes, U-Net keeps a clear SSIM lead over CLASSIC, comparable to the gap already seen at single-pass. On LPIPS the two models are close pooled,
down sharply from the wide single-pass gap. This may look like LPIPS has become insensitive to rollout length but it has not. We see that  Figure~\ref{fig:ark50_bar_overall} shows that U-Net's LPIPS is worse than CLASSIC's specifically in the high/high regime and better than
CLASSIC's in the other three. The pooled average cancels these two opposite effects in the same regime-dependent way already documented above for PS log-MSE.

We caution against reading U-Net's remaining lead on NRMSE(mean/std) and SSIM as evidence that it is the better model overall. These
are RMSE-based metrics minimised by the conditional mean of the predictive distribution, which is what U-Net is trained to output but the CLASSIC/NOCOND samples are not. The metrics are therefore biased.
\citet{saharia2022} make the same point explicitly when justifying their choice of evaluation metrics for image-to-image diffusion models, avoiding PSNR/SSIM as measures of sample quality for tasks that require hallucination and noting that these metrics are known to
prefer blurry regression outputs over perceptually accurate ones \citep{dahl2017,ledig2017,menon2020}. The metrics least susceptible to this bias, LPIPS and PS log-MSE, are precisely the ones where CLASSIC's
advantage is largest and grows with both dynamism and rollout length.

\begin{figure}
    \centering
    \includegraphics[width=\linewidth]{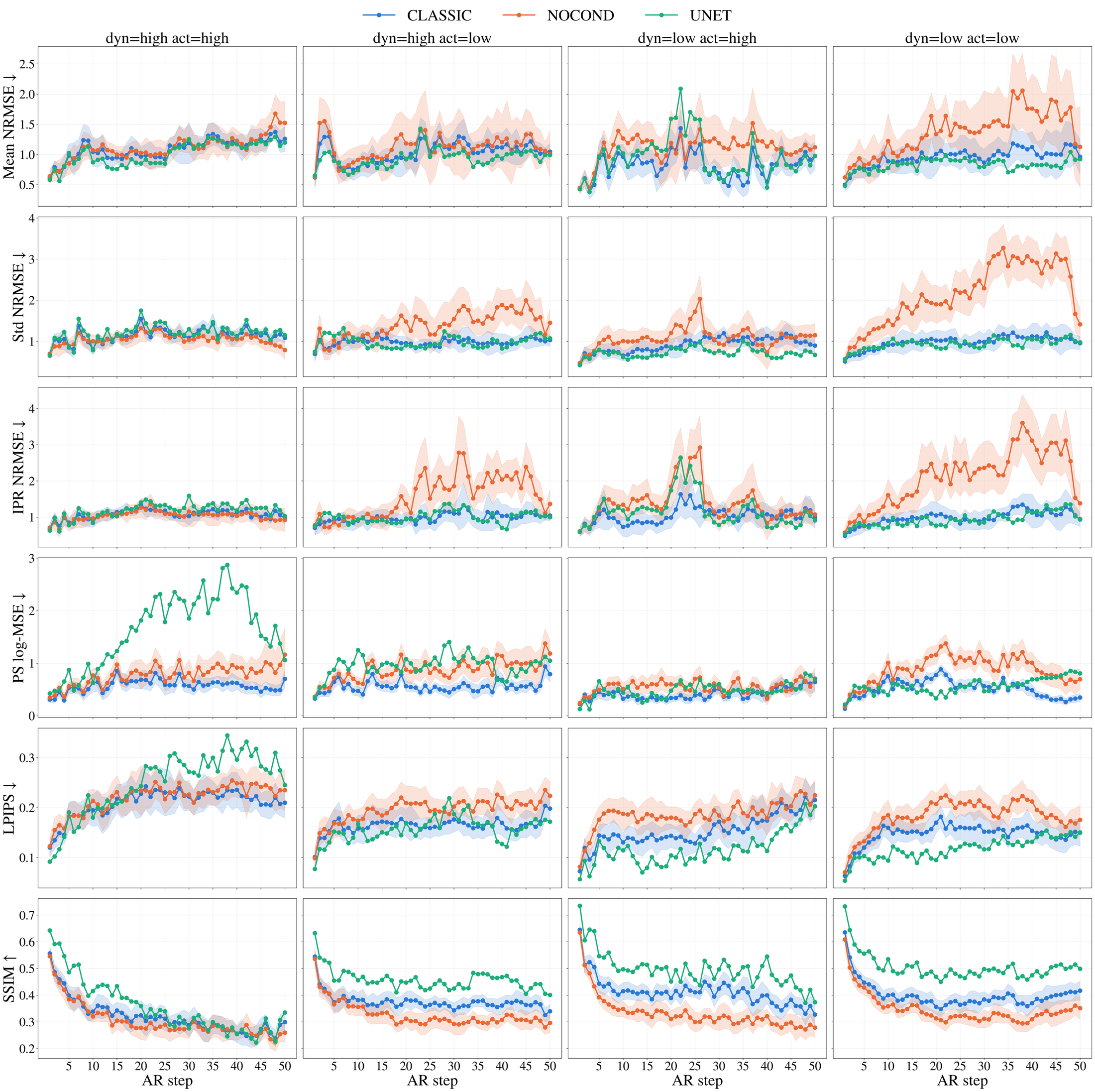}
    \caption{Evolution of NRMSE(mean/std/IPR) and PS log-MSE/LPIPS/SSIM across the 50 AR steps (350
    frames, 11h40), by activity$\times$dynamism regime (columns). Each point averages over the 7 frames of
    that AR step and over the matched sequences of Figure~\ref{fig:ark50_nrmse_overall}.}
    \label{fig:ark50_degradation_by_ar_step}
\end{figure}

\textbf{Per-step degradation.} Figure~\ref{fig:ark50_degradation_by_ar_step} shows that U-Net's PS log-MSE and LPIPS degrade steeply with AR step count, but \emph{only in the high-activity/high-dynamism regime}: PS log-MSE peaks at more than four and a half times CLASSIC's value around step 38, before partially easing by step 50. LPIPS shows an analogous pattern but with the sign reversed. U-Net is better than CLASSIC at step 1, crosses over to worse by step 5, and reaches its own worst point at step 38. SSIM mirrors this: U-Net outperforms at step 1 and stays ahead until roughly step 25--30, after which the two curves interleave for the remainder of the rollout, no longer showing a consistent U-Net advantage.
In the other three regimes, by contrast, U-Net's PS log-MSE, LPIPS, and SSIM curves stay close to or better than CLASSIC's for most or all of the 50 steps (consistent with U-Net's regime-level wins in Figure~\ref{fig:ark50_bar_overall}). The mechanism is the same regression-to-the-mean bias discussed throughout this section, but it scales with how much genuine high-frequency structure is available to lose. In a calm, low-activity sequence, a smoothed trajectory forfeits little. Whereas in a highly dynamic, active one, the same smoothing bias becomes the dominant source of error as it compounds over many time-steps. This is a central result to this study and a single-pass evaluation, limited to a 7-frame window is not able to find this. 

\begin{figure}
    \centering
    \includegraphics[width=\linewidth]{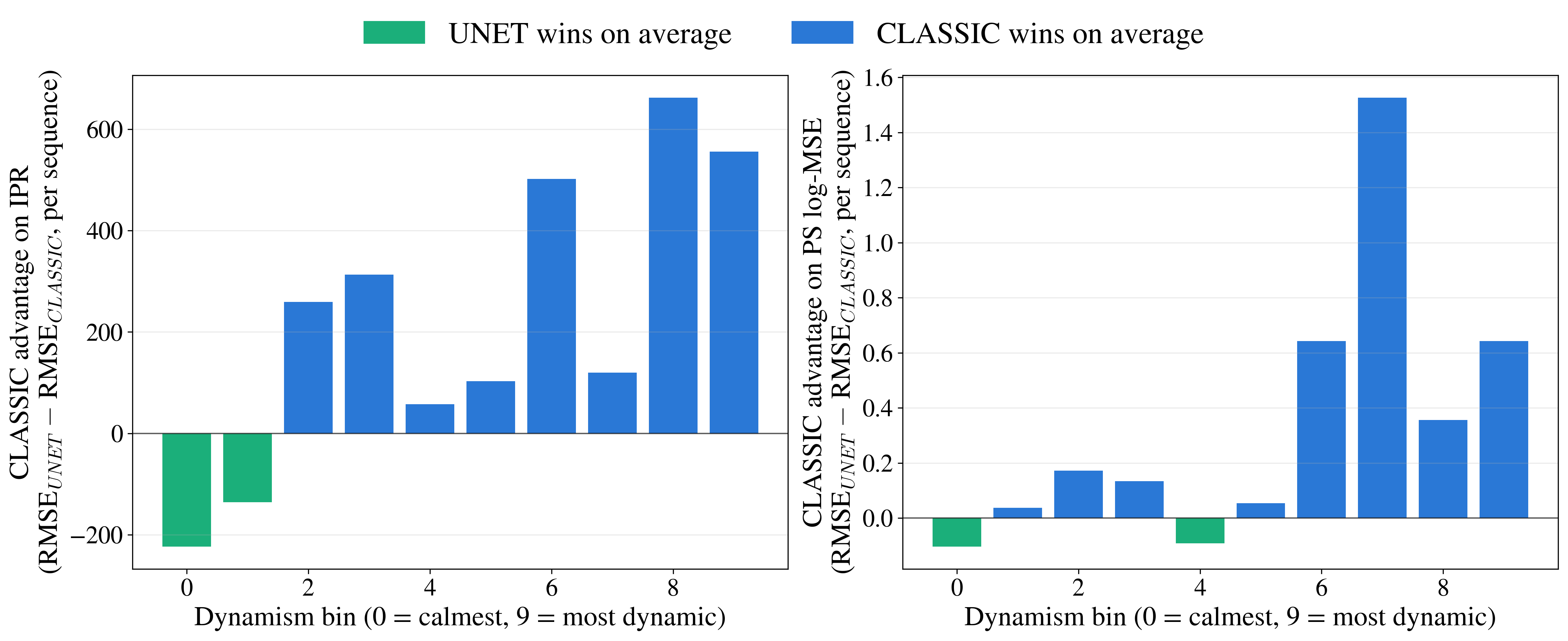}
    \caption{CLASSIC's advantage over U-Net at $K=50$, per sequence, on IPR and PS log-MSE
    (RMSE$_{\text{U-Net}}$ $-$ RMSE$_{\text{CLASSIC}}$, each computed over the full 350-frame rollout;
    positive favours CLASSIC), binned by decile of true per-sequence dynamism and pooled across all four
    regimes ($n=62$ sequences). Same construction as Figure~\ref{fig:singlepass_dynamism_decile_all}, at
    $K=50$ instead of single-pass. Bars are coloured by which model wins on average in that decile.}
    \label{fig:ark50_dynamism_decile_all}
\end{figure}

\textbf{Dynamism dependence.} At single-pass
(Figure~\ref{fig:singlepass_dynamism_decile_all}), CLASSIC only overtakes U-Net in the top two or three deciles (from roughly decile 7--8 onward on IPR and PS log-MSE respectively). At $K=50$ (Figure~\ref{fig:ark50_dynamism_decile_all}) CLASSIC performs best from the second-calmest decile upward, i.e.\ across roughly 80\% of
the dynamism range instead of 20\%. CLASSIC's advantage is also the magnitude dynamics which is roughly an order of magnitude larger at $K=50$ than at single-pass, on both IPR and PS log-MSE (Figure~\ref{fig:ark50_dynamism_decile_all}). CLASSIC's stochastic diversity therefore becomes advantageous across a much wider share of the dynamism spectrum once the rollout is long enough for it to compound. A single 14-minute window bounds how much this effect can accumulate but 11h40 permits substantially more accumulation.

\subsection{Out-of-Distribution Case Study: The 2015 St.\ Patrick's Day Storm}
\label{subsec:results_2015}

We now push the autoregressive rollout far beyond the 11h40 of Section~\ref{subsec:results_fakear}, to the multi-day, fully out-of-distribution event described in Section~\ref{subsec:st_patrick_test}
(Figure~\ref{fig:st_patrick_event}). We generate from three \texttt{PRED\_START} windows chosen to sample distinct phases of the storm, 16 March 04:34 and 19:52 UT (pre-storm quiet baseline) and 17 March 06:00 UT (main-phase onset, Figure~\ref{fig:st_patrick_event}), and roll out every window to a common end time of 19 March 00:00 UT, yielding trajectories between 1260 and 2023 predicted frames
(42.0--67.4 hours). As in Section~\ref{subsec:results_fakear}, only the potential map is fed back auto-regressively and the upstream L1 driver remains observed ground truth throughout. Unlike the activity/dynamism-stratified evaluations of Sections~\ref{subsec:results_singlepass} and~\ref{subsec:results_fakear}, here we report a single intense event and include the persistence baseline throughout, which becomes competitive over a multi-day horizon.

\begin{figure}
    \centering
    \includegraphics[width=\linewidth]{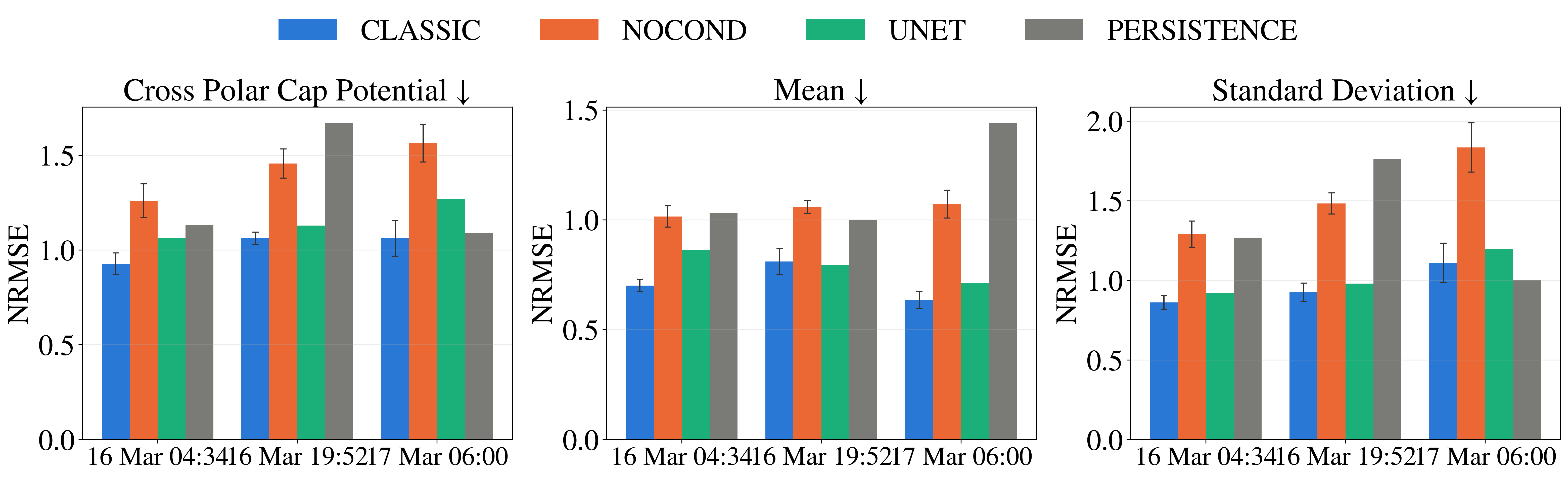}
    \caption{NRMSE of the mean potential, spatial standard deviation, and IPR, averaged over the full
    rollout (42.0--67.4 hours) of each of the three \texttt{PRED\_START} windows.}
    \label{fig:storm2015_nrmse_overall}
\end{figure}

\begin{figure}
    \centering
    \includegraphics[width=\linewidth]{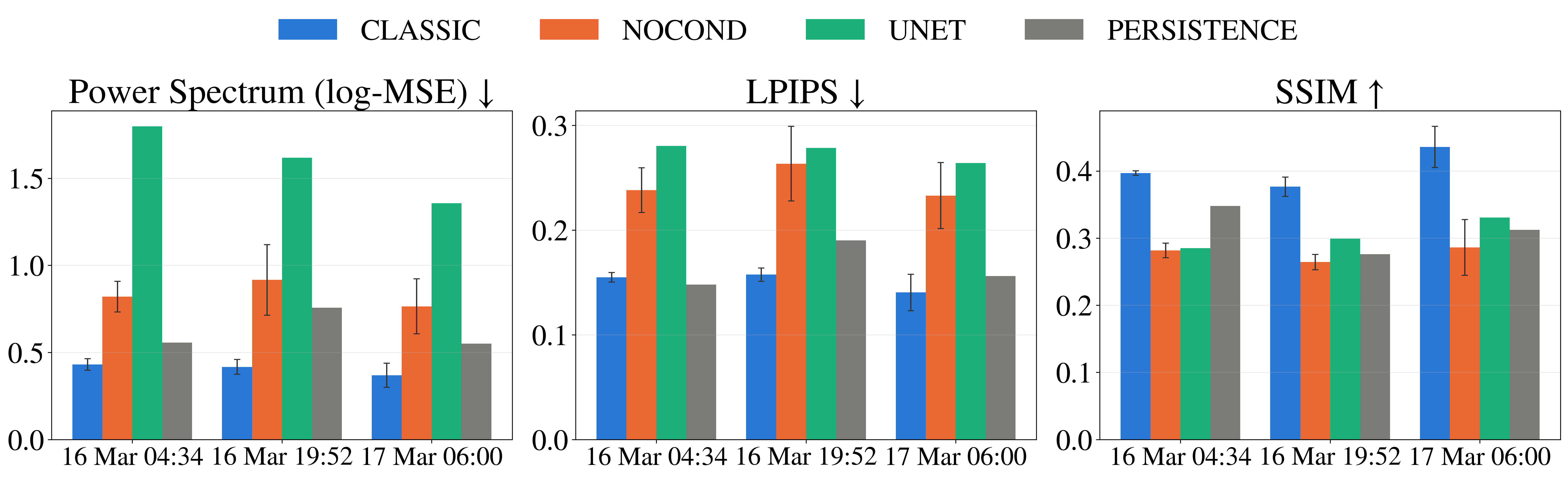}
    \caption{PS log-MSE, LPIPS, and SSIM, averaged over the full rollout of each of the three
    \texttt{PRED\_START} windows.}
    \label{fig:storm2015_bar_overall}
\end{figure}

\textbf{Global scalar statistics.} At single-pass (Section~\ref{subsec:results_singlepass}), NRMSE(mean/std/IPR) showed limited power to discriminate between models, with differences of only a few percent. Here, under the longer rollout, clear trends emerge. CLASSIC attains the lowest NRMSE on IPR (1.02 vs.\ 1.15 for U-Net and 1.43 for NOCOND) and mean (0.72 vs.\ 0.79 for U-Net and 1.05 for NOCOND), and also leads, more narrowly, on the standard deviation NRMSE (0.97 vs.\ 1.03 for U-Net) (Figure~\ref{fig:storm2015_nrmse_overall}). U-Net remains competitive here, occasionally matching or marginally beating CLASSIC on individual windows for the mean, consistent with the mechanism identified at single-pass and $K=50$. NOCOND is the weakest model on every scalar statistic in every window, with NRMSE(std) reaching 1.83, worse than predicting a constant.

\textbf{Structural and spectral metrics.} Averaged across the three windows, CLASSIC attains the lowest PS log-MSE and LPIPS, and, for the first time in this evaluation, the highest SSIM (Figure~\ref{fig:storm2015_bar_overall}). This contradicts the finding of Section~\ref{subsec:results_singlepass}, where U-Net led on both LPIPS and SSIM, and extends the partial crossover seen at $K=50$ (Section~\ref{subsec:results_fakear}), where CLASSIC had
overtaken U-Net on PS log-MSE while U-Net outperforms on LPIPS and SSIM. Over a multi-day rollout, U-Net's regression-to-the-mean bias is compounded at every autoregressive steps, and its structural similarity to the ground truth eventually degrades below even the diffusion models' individual stochastic samples. Notably, CLASSIC is also the only model to consistently beat the persistence baseline on all three perceptual/spectral metrics, whereas NOCOND and U-Net are both beaten by the persistence on PS log-MSE and LPIPS in every window
(Figure~\ref{fig:storm2015_bar_overall}).

\begin{figure}
    \centering
    \includegraphics[width=\linewidth]{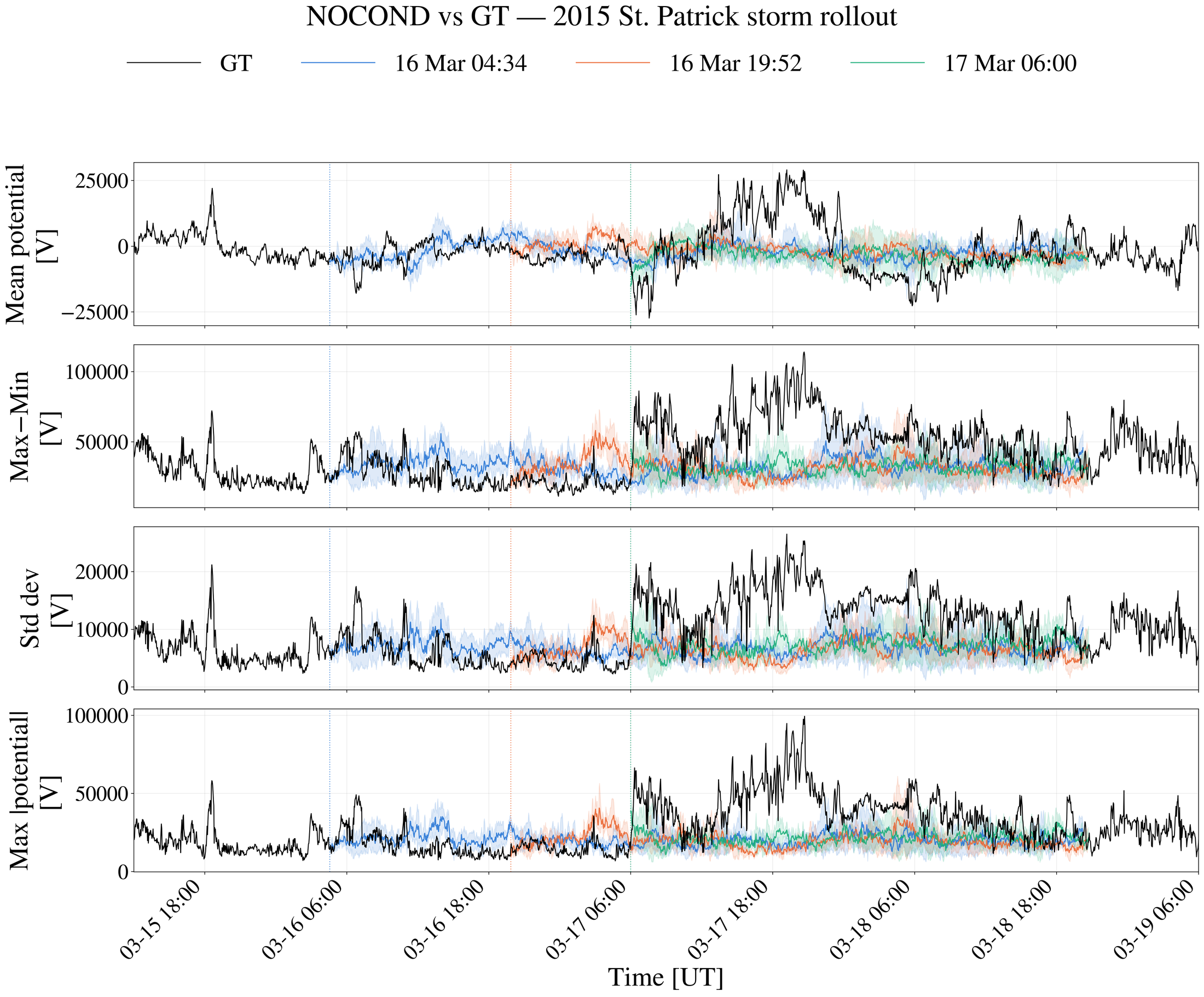}
    \caption{NOCOND (no L1 conditioning) vs.\ ground truth, per-frame mean/max$-$min/std/max$|\Phi|$, for each of the three \texttt{PRED\_START} windows.}
    \label{fig:storm2015_nocond}
\end{figure}

\begin{figure}
    \centering
    \includegraphics[width=\linewidth]{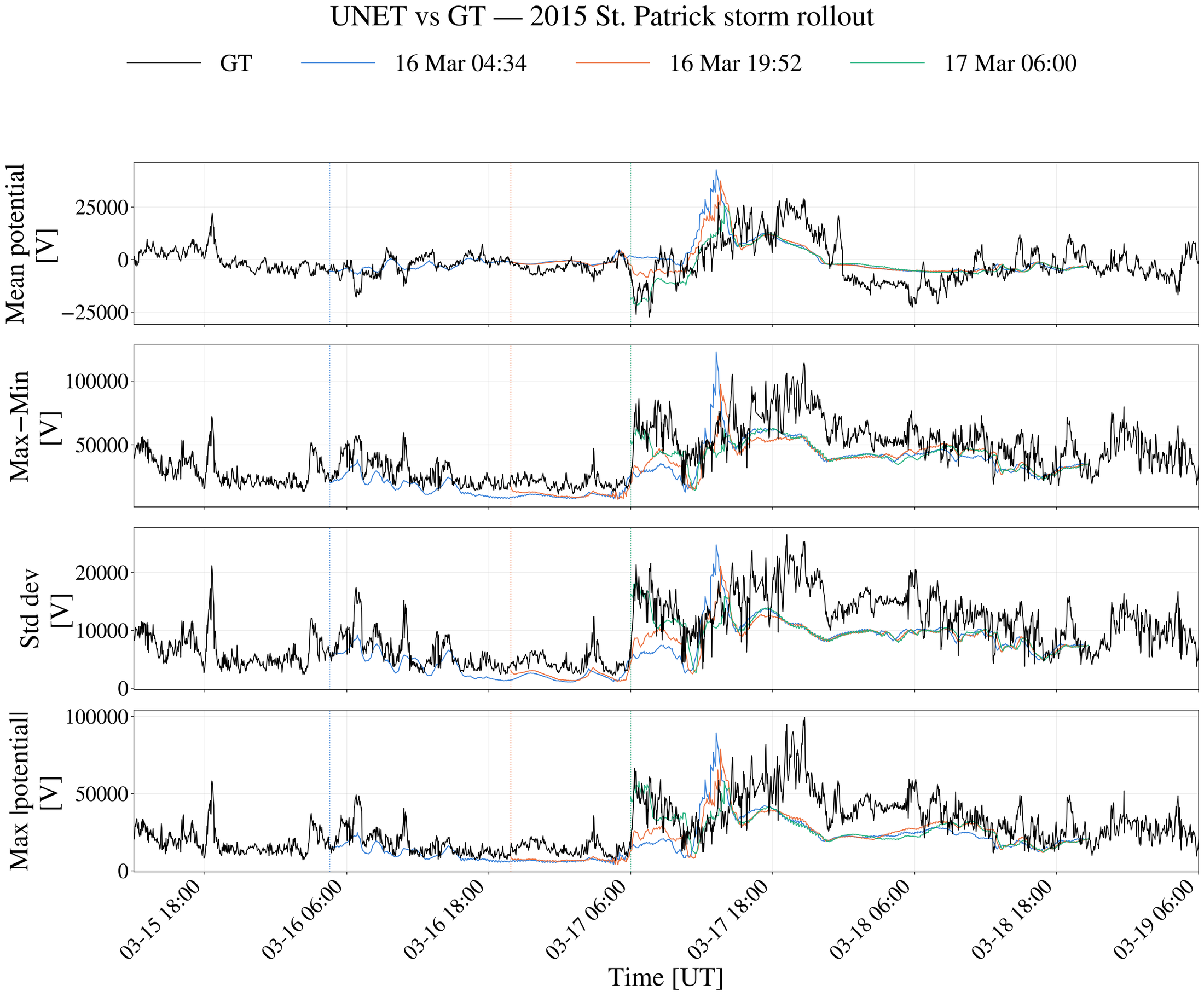}
    \caption{U-Net vs.\ ground truth, same layout as Figure~\ref{fig:storm2015_nocond}.}
    \label{fig:storm2015_unet}
\end{figure}

\begin{figure}
    \centering
    \includegraphics[width=\linewidth]{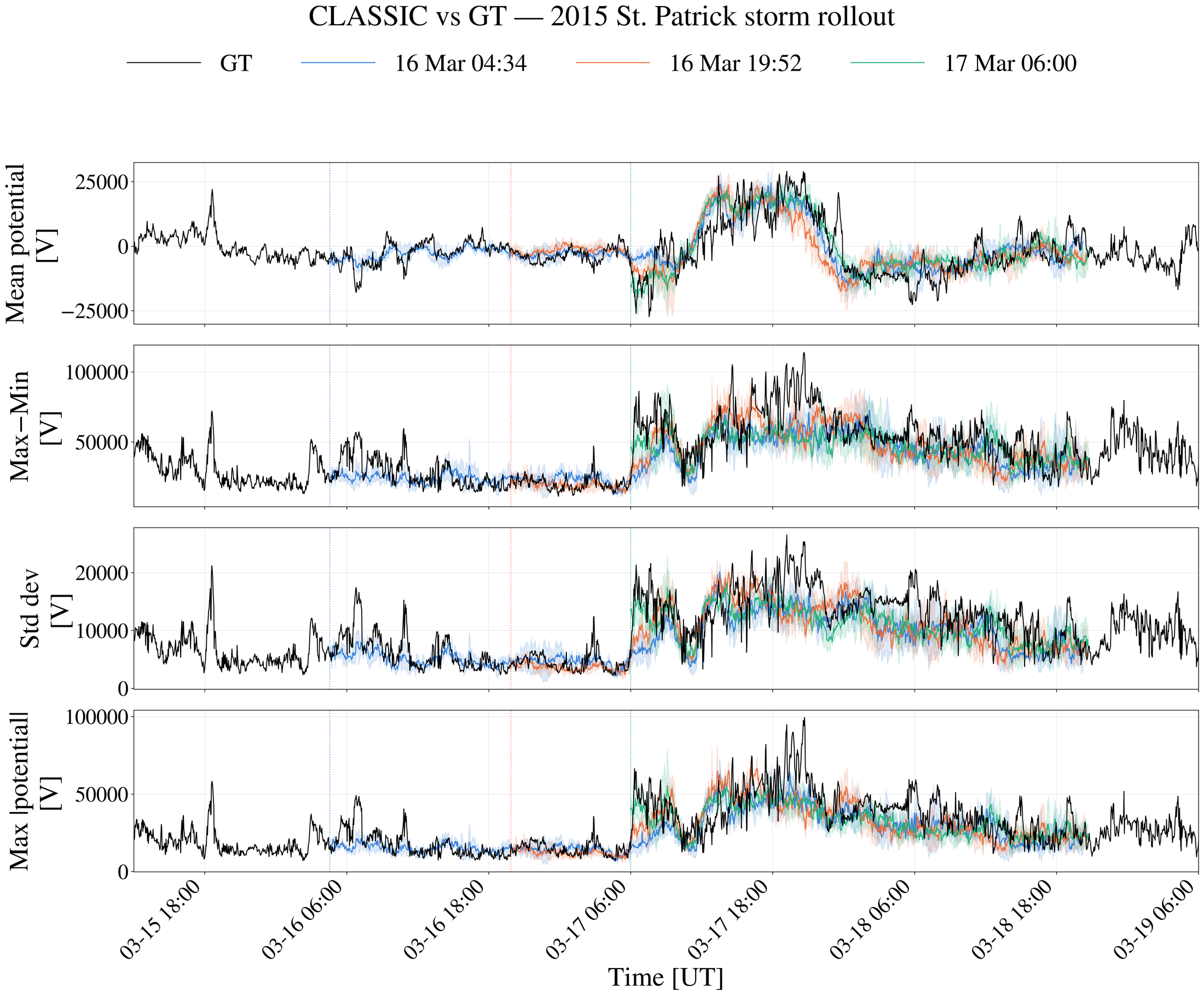}
    \caption{CLASSIC vs.\ ground truth, same layout as Figure~\ref{fig:storm2015_nocond}. Shaded bands
    denote $\pm 1$ standard deviation across the 3 diffusion samples.}
    \label{fig:storm2015_classic}
\end{figure}

\textbf{Per-window trajectories.}
NOCOND (Figure~\ref{fig:storm2015_nocond}), lacking any conditioning on the upstream L1 driver, tracks the ground truth closely during the two pre-storm windows but is entirely blind to the main-phase
onset in the third window. We see this in the mean, max$-$min, standard deviation, and max$|\Phi|$ which remain at pre-storm levels for the full rollout while the ground truth's max$-$min triples from $\sim$20--30~kV to 80--110~kV. Without access to the solar wind driver, the model has no information on which to anticipate the intensification. U-Net (Figure~\ref{fig:storm2015_unet}), which does retain conditioning, reacts accordingly. All three windows show a jump synchronised with the ground truth immediately after 17 March 06:00 and the window starting 16 March 04:34 follows the shape of the amplitude peak between 17 and 18 March. The deterministic model's advantage is short-lived as the same compounding regression-to-the-mean
mechanism established in Section~\ref{subsec:results_fakear} takes hold. By 12--18 hours past onset the predicted amplitude has become an over-smoothed trajectory, losing the high-frequency variability the ground truth sustains through 19 March. CLASSIC (Figure~\ref{fig:storm2015_classic}) combines both positive qualities as it reacts at onset like U-Net, but its per-step stochastic sampling regenerates plausible high-frequency structure at every autoregressive
step. As a result, it tracks the ground truth through both the onset and the multi-day recovery, remaining within $\pm 1$ sample standard deviation for most of the $\sim$2.5-day rollout.

\begin{figure}
    \centering
    \includegraphics[width=\linewidth]{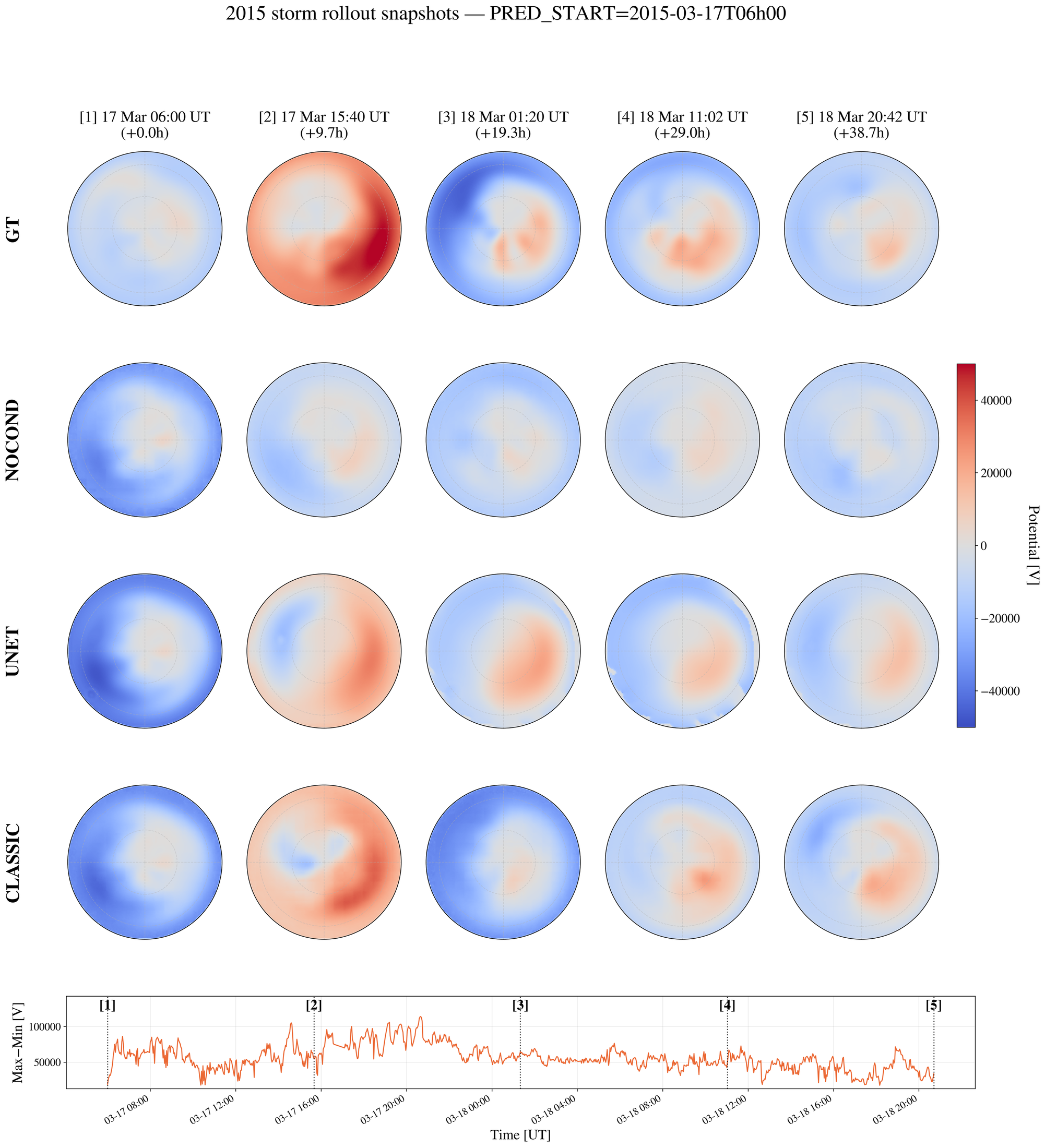}
    \caption{Ground truth and predicted potential maps (17 March 06:00 window; CLASSIC/NOCOND sample~0)
    at five instants evenly spaced across the full rollout. The ground-truth max$-$min timeseries below
    marks where each instant falls (numbered dotted lines).}
    \label{fig:storm2015_frames_evenly}
\end{figure}

\begin{figure}
    \centering
    \includegraphics[width=\linewidth]{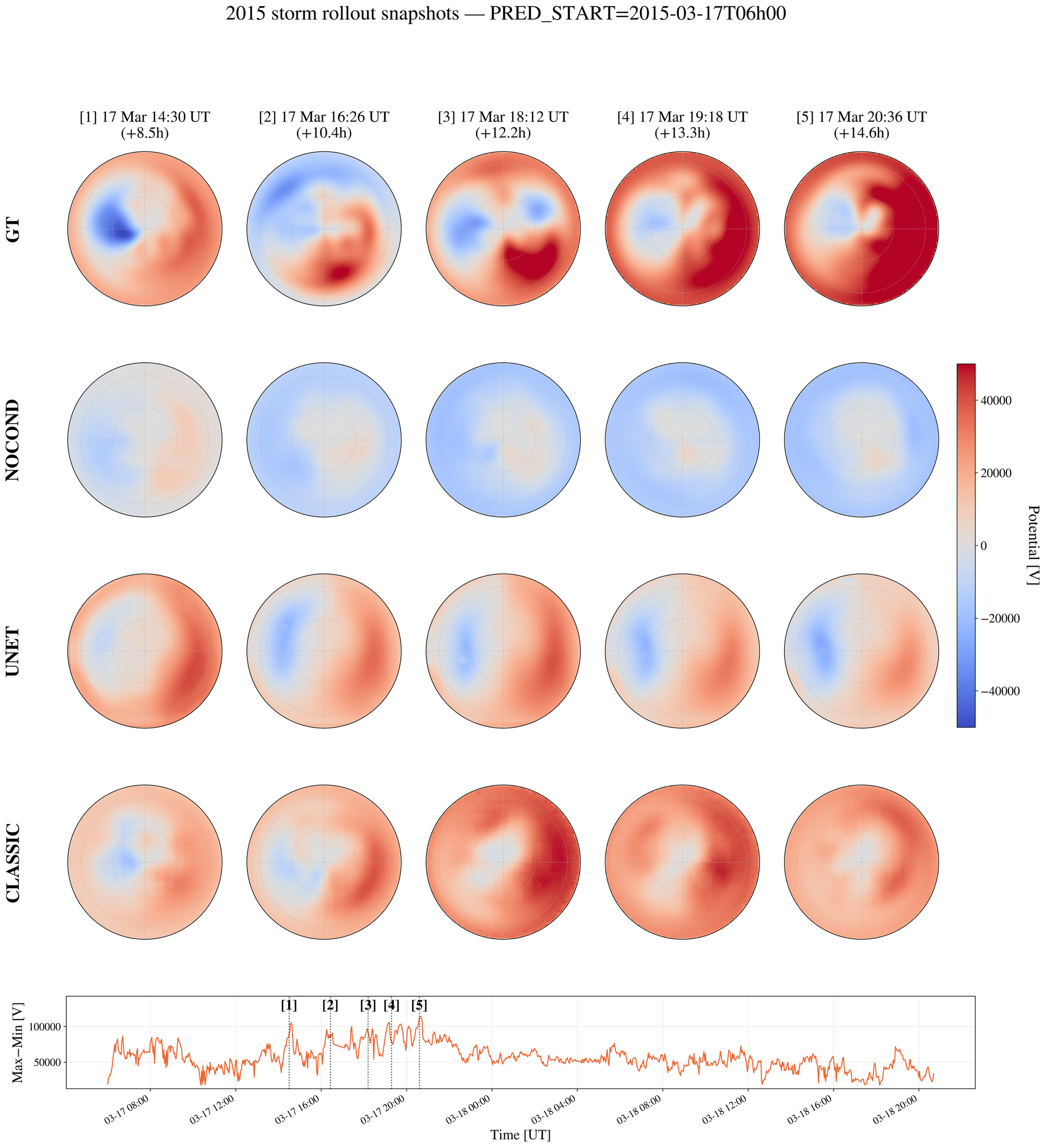}
    \caption{Same layout as Figure~\ref{fig:storm2015_frames_evenly}, zoomed into the five most
    prominent local peaks of the ground-truth max$-$min curve during the main-phase build-up
    (8.5--14.6 hours after \texttt{PRED\_START}).}
    \label{fig:storm2015_frames_peaks}
\end{figure}

In Figure~\ref{fig:storm2015_frames_evenly} shows examples from thequieter parts of the storm. The NOCOND maps resemble the pale, pre-storm pattern of instant~1 even at instant~2, when the ground truth is saturated with the storm's intense positive potential. U-Net and CLASSIC both capture the reddening (increase in positive potential) at instant~2 and track the localised positive patch that persists through recovery (instants~3--5), though U-Net's spatial pattern is visibly smoother and less textured than either the ground truth or CLASSIC's sample throughout. Figure~\ref{fig:storm2015_frames_peaks} zooms into the extreme part of the storm. The ground truth's negative-potential region on the dawn side visibly shrinks while its positive potential cell expands and rotates across the five frames. U-Net reproduces the growing intensity, but its negative-potential feature stays close to fixed in shape and position across all five snapshots, consistent with the frozen, over-smoothed spatial structure implied by its rising PS log-MSE/LPIPS in Figure~\ref{fig:ark50_degradation_by_ar_step}, while CLASSIC's negative region visibly shrinks and its positive cell's asymmetric shape evolves in a way that more closely tracks the ground truth. We stress that this is a single sample from a single window, offered as a qualitative illustration rather than additional quantitative evidence.

The three evaluations of Section~\ref{sec:results} reveal that U-Net's regression-to-the-mean bias favours means it performs best at the short, single-pass horizon, but this advantage narrows by $K=50$. Here, CLASSIC overtakes U-Net on PS log-MSE, and CLASSIC leads on every metric at the multi-day, out-of-distribution storm. The single-pass evaluation underestimates CLASSIC's long-horizon autoregressive advantage.

These results should be interpreted with several caveats in mind. First, the ground-truth electric potential maps are themselves not raw observations but a spherical harmonic fit to the observed line-of-sight velocity vectors, supplemented with the \citet{cousins2010} climatological background wherever radar backscatter is absent (Section~\ref{subsec:superdarn}). The models therefore learn to reproduce a processed, partially model-constrained product, whose own uncertainty grows in regions and periods of sparse SuperDARN coverage. Second, the analysis is restricted to the Southern Hemisphere, chosen because it is the more data-sparse and geometrically complex of the two hemispheres (Section~\ref{subsec:superdarn}). Whether the same crossover in relative model skill holds under the denser Northern Hemisphere coverage remains untested. Finally, all three evaluation regimes, are drawn from the moderate range of Solar Cycle 25 and the single most extreme event available from Solar Cycle 24. Unprecedented solar wind conditions, could expose failure not captured by the present analysis.

\section{Conclusion}

We presented a ViT-based diffusion model for forecasting high-latitude ionospheric convection, conditioned on multi-variate solar wind and IMF observations at L1, and benchmarked it against a deterministic U-Net baseline and an unconditioned diffusion ablation (NOCOND) across four progressively more demanding evaluation regimes. Our central finding is that the relative merit of the deterministic and probabilistic approaches is not fixed but shifts systematically with the length of the autoregressive rollout and, within any given horizon, with the true dynamism of the underlying event. U-Net's regression-to-the-mean bias is advantageous over short horizons and calm convection, but compounds errors over long, dynamic rollouts.

The unconditioned ablation clarifies that this generalisation is not solely a property of the diffusion objective. NOCOND, which shares CLASSIC's architecture and stochastic sampling but has no access to the L1 solar wind driver, is blind to the 2015 storm's onset, remaining pinned at pre-storm amplitude for the full rollout while the true field's dynamic range triples. Skill under out-of-distribution forcing therefore requires both the conditioning signal and an architecture that is able to keep exploiting the signal without collapsing to a smoothed state over long autoregressive horizons.

The out-of-distribution evaluation is a single intense storm. It was chosen to be extreme along multiple axes (different solar cycle, different L1 instrument, longest tested horizon), but a single event cannot establish how these findings generalise across the full space of possible storms. The autoregressive rollout in every test remains driven by the true, observed L1 timeseries, so the reported horizons are not yet a fully autonomous forecast.

These results argue for probabilistic, sample-based generative models as the more promising direction for operational ionospheric forecasting, because operational value concentrates at the long horizons and extreme, out-of-distribution events where deterministic regression degrades most. A natural next step is coupling this model with an upstream module that forecasts L1 solar wind and IMF conditions from solar disk observations, removing the dependence on an observed upstream driver and enabling a fully end-to-end Sun-to-Earth forecasting pipeline (Section~\ref{subsubsec:single_pass_vs_autoregressive}). Extending the evaluation to include ionospheric flow velocities and out-of-distribution evaluation to additional historical storms spanning other solar cycles, and to the Northern Hemisphere, would further test the generalisation advantage identified here.

\section*{Open Research Section}
The code, trained model configurations, and the paired SuperDARN--L1 dataset (including the ballistic-pairing and deduplication scripts) used in this study are openly available on GitHub at \url{https://github.com/fpramunno/ionosphere_diffusion}. A permanent, versioned archive of the code and dataset will additionally be deposited on Zenodo prior to publication.

\section*{Conflict of Interest disclosure}
The authors declare there are no conflicts of interest for this manuscript.

\acknowledgments
The authors acknowledge the use of SuperDARN data. SuperDARN is a collection of radars funded by national scientific funding agencies of Australia, Canada, China, France, Italy, Japan, Norway, South Africa, United Kingdom, and the United States of America. The SuperDARN "rawacf" data used in this study have been retrieved from the repository at British Antarctic Survey, upon official request and dedicated account (\url{https://www.bas.ac.uk/project/superdarn/}).
This work was supported by the Italian National Programme for Antarctic Research (PNRA) through its contract n. PNRA 2022$\_$0000086.
This work was supported by the Swiss National Supercomputing Center (CSCS) with a grant from for the Swiss SKA Consortium under project ID sk035. MTW acknowledges funding from UKRI STFC through the Ernest Rutherford Fellowship (ST/X003663/1).

%
%

\bibliography{references}

%
%
%
%
%

\end{document}